\documentclass[twocolumn]{aastex63}

\usepackage{amsmath}

\newcommand{\gcc}{\mathrm{g~cm^{-3}}}

\newcommand{\Lthin}{\mathcal{L}_\mathrm{thin}}
\newcommand{\ffa}{\mathrm{ff}}
\newcommand{\ssa}{\mathrm{ssa}}
\newcommand{\csm}{\mathrm{csm}}
\newcommand{\timp}{t_\mathrm{imp}}

\newcommand{\rhocsm}{\rho_\mathrm{csm}}
\newcommand{\Rin}{R_\mathrm{in}}
\newcommand{\DeltaR}{\widetilde{\Delta r}}
\newcommand{\ugas}{u_\mathrm{gas}}

\received{January 29, 2021}
\accepted{\today}
\submitjournal{ApJ}

\shorttitle{SN\,Ia Shell CSM}
\shortauthors{Harris et al.}

\begin{document}

\title{Tumbling Dice: Radio Constraints on the Presence of Circumstellar Shells around Type Ia Supernovae with Impact Near Maximum Light}

\author[0000-0002-1751-7474]{Chelsea E.~Harris}
\affiliation{Center for Data Intensive and Time Domain Astronomy, Department of Physics and Astronomy, Michigan State University, East Lansing, MI 48824, USA}

\author[0000-0002-8400-3705]{Laura Chomiuk}
\affiliation{Center for Data Intensive and Time Domain Astronomy, Department of Physics and Astronomy, Michigan State University, East Lansing, MI 48824, USA}

\author[0000-0002-3389-0586]{Peter. E.~Nugent}
\affiliation{Lawrence Berkeley National Laboratory, 1 Cyclotron Road, MS 50B-4206, Berkeley, CA 94720, USA}
\affiliation{Department of Astronomy, University of California, Berkeley, CA 94720-3411, USA}

\correspondingauthor{CEH}
\email{harr1561@msu.edu}

\begin{abstract}

The progenitors of Type Ia supernovae (SNe\,Ia) are 
debated, particularly the evolutionary state of the binary companion
that donates mass to the exploding carbon-oxygen white dwarf.
In previous work, we presented hydrodynamic models and 
optically thin radio synchrotron light-curves 
of SNe\,Ia interacting with detached, confined shells of CSM, 
representing CSM shaped by novae.
In this work, we extend these light-curves to the optically thick
regime, considering both synchrotron self-absorption and free-free 
absorption. 
We obtain simple formulae to describe the evolution of optical
depth seen in the simulations, allowing optically thick light-curves
to be approximated for arbitrary shell properties. 
We then demonstrate the use of this tool by interpreting published radio data.
First, we consider the non-detection of PTF11kx -- 
an SN\,Ia known to have a detached, confined shell --
and find that the non-detection is consistent with
current models for its CSM, and that observations at a later 
time would have been useful for this event.
Secondly, we statistically analyze an ensemble of radio non-detections 
for SNe\,Ia with no signatures of interaction, 
and find that shells with masses $(10^{-4}-0.3)~M_\odot$ 
located $(10^{15}-10^{16})~\mathrm{cm}$ from the progenitor are currently
not well constrained by radio datasets, due to their dim, rapidly-evolving
light-curves.
\end{abstract}


\section{Introduction} \label{sec:intro}

Thermonuclear Type Ia supernovae (SNe\,Ia) are one of the most mature and precise cosmological tools in modern astronomy, and have revealed the accelerating expansion of the universe \citep{Riess+98,Perlmutter+99}.
SNe\,Ia are the explosion of a carbon-oxygen white dwarf that has merged with or accreted mass from a companion star. 
However, we 
     
    currently remain ignorant of the identity of the companion star, 
    which
    affects the timescale of explosion, explosion trigger, 
    properties of the white dwarf at time of explosion, and local environment. 
    
    It has long been recognized that 
    characterizing the circumstellar material (CSM) around SNe\,Ia 
    constrains the nature of their companions \citep{Branch+95}.
    For example, main sequence and red giant companions
    of the ``single-degenerate'' channel will have winds.
    Growth of the white dwarf happens through accretion, either
    Roche-lobe overflow or directly from the companion wind
    (i.e., a symbiotic system).
    Instabilities in this mass-transfer (e.g., novae) can create 
    dense shells of hydrogen-rich CSM. 
    In contrast, the ``double-degenerate'' channel, where explosion
    is triggered by the merger of two white dwarfs, is expected
    to have a clean environment.

    Radio observations are sensitive probes of the CSM around SNe, 
    as synchrotron emission is produced when the SN blast wave shocks 
    surrounding gas, accelerates electrons to relativistic speeds, 
    and amplifies the magnetic field in the shocked region \citep{Chevalier82}.      
    This emission will be subject to absorption, and of particular relevance
    to this work is the absorption caused by the CSM itself: 
    within the shock region, radio emission is affected by synchrotron self-absorption,
    and radio emission emerging from the shock region is further subject
    to free-free absorption from outlying, unshocked CSM. 
    Despite extensive observations of SNe\,Ia at radio wavelengths, 
    there are no published radio detections of SNe\,Ia to date, 
    even for those known to be interacting. 
    Upper limits on radio luminosity imply that the CSM around typical 
    SNe\,Ia is substantially lower density than observed around most core-collapse SNe,
    assuming the CSM is a continuous medium like a wind
    \citep{Weiler+02,Perez-Torres+14,Chomiuk+16,Lundqvist+20}.

    In recent decades, the picture of SN~Ia environments and the single-degenerate
    channel has become muddied by the discovery of what was long-sought:
    SNe~Ia with signatures of hydrogen (from CSM interaction) in their spectra, 
    dubbed SNe~Ia-CSM by \citet{Silverman+13}. 
    SNe~Ia-CSM can be broken into two groups.
    The first and most common 
    are events like SN~2005gj, which were historically grouped with
    the canonical CSM interaction class of SNe~IIn but have distinct underlying
    SN~Ia features.
    Radio non-detections are expected for such events 
    since light at radio frequencies will be totally absorbed 
    by the outlying CSM that has not yet been shocked. 
    The other case is more rare, where an SN~Ia transforms from a normal event
    into an interacting event \citep[which we for shorthand call SNe~Ia;n,][]{Harris+18}.
    The prototype is PTF11kx \citep{Dilday+12}, 
    though SN~2002ic may also have been an SN~Ia;n \citep{WoodVasey+04}.
    A search for more instances of SNe~Ia;n discovered interaction in
    SN~2015cp \citep{Graham+19}. 
    The CSM of SNe~Ia;n may be shaped by nova outbursts or other 
    instabilities in the mass-transfer process that sweep any existing material
    into a distant shell.

    SNe~Ia;n are of particular interest because they are so disruptive to
    our current theoretical understanding and abilities, 
    and to SN~Ia observational traditions --- yet they are a clear
    path forward to understanding the single-degenerate channel.
    They disrupt our theoretical understanding because novae should 
    interrupt the mass growth of the carbon-oxygen white dwarf 
    \citep[see, e.g., the discussion and references in][]{Branch+95}.
    Yet the CSM mass observed for PTF11kx \citep{Graham+17} was too low
    to have come from an expelled common envelope of the double-degenerate 
    scenario \citep[][]{LivioRiess03}. 
    They disrupt theoretical ability because the well-established
    tools for interpreting interaction with a wind or other continuous medium
    cannot be applied \citep{Chevalier82}. 
    SNe~Ia;n are furthermore extremely difficult to detect via traditional
    SN~Ia observation methods, which only cover the phase near maximum light
    during which the CSM will not be visible (even in spectra, 
    as was the case for SN~2015cp). 
    Furthermore, the interaction may be very short-lived,
    eluding even observations at late times --- and the fact that the time
    between mass ejection and supernova is unknown means the location of
    the shell is unknown and potentially random.
    Finally, as is the case with the SN~IIn-like events, SNe~Ia;n 
    come from the rare ``shallow silicon'' or ``SN~1991T-like'' subgroup
    of SNe~Ia, making the chance of discovery even smaller since SN~Ia surveys
    usually attempt to recreate the underlying distribution of SN~Ia properties. 
    Despite all of these difficulties, SNe~Ia;n are the clearest path forward
    to understanding the single-degenerate channel, because we can constrain
    the ejecta properties from pre-interaction data to alleviate degeneracies
    in the interaction modelling, and because the CSM mass is too low to be
    explained by a double-degenerate origin (as aforementioned).

    The potential for SNe~Ia;n to illuminate SN~Ia progenitors motivates
    the alleviation of the theoretical obstacles facing their study.
    \citet[][Paper~I]{HNK16} modeled SNe\,Ia interacting with low-mass,
    confined shells of CSM in the months following maximum light. 
    The optically thin radio light-curves from these models were then studied, 
    and a parameterization was created to allow for light-curves to 
    be created for an arbitrary CSM shell configuration. 
    These light-curves can be used to limit CSM shell properties
    from radio non-detections particularly for the very thin, low-mass
    shells expected from single nova eruptions, distant shells that will
    be very low density, or radio observations taken after interaction has ended
    \citep{Harris+18,Cendes+20, Pellegrino+20}. 

    However, there is a sizeable sample of SNe~Ia with radio 
    observations near maximum light, probing shells at a distance
    $r\sim(10^{15}-10^{16})~\mathrm{cm}$ \citep{Chomiuk+16},
    and the use of this dataset is currently limited by the optically
    thin assumption, which is only applicable to shells of density
    $\lesssim10^{-17}~\gcc$ or after the shock has crossed
    the shell.
    In order to study interaction within $r\sim10^{16}~\mathrm{cm}$ and
    to incorporate lower-frequency observations, the optically thin 
    light-curves of Paper~I
    must be extended into the regime of synchrotron self-absorption
    and external free-free absorption,
    which is our aim for this work.
    With absorption accounted for, we can use the radio sample to 
    constrain the presence of nova-like shells around SNe~Ia for higher
    shell masses than was previously possible. 

    This work is organized as follows. 
    In \S~\ref{sec:PaperI}, we summarize the main results of Paper~I
    for the reader's convenience. 
    We then present the method for modifying the optically thin
    luminosity by the photon escape fraction to obtain a light-curve
    with absorption in \S~\ref{sec:lum_calc}.
    We account for synchrotron self-absorption ($\tau_\ssa$) 
    and free-free ($\tau_\ffa$) absorption.
    In \S~\ref{sec:tau_calc} we show the evolution of optical depth
    as calculated directly from the hydrodynamic models.
    The creation of optically-thick light-curves for an arbitrary 
    shell configuration without the need for hydrodynamic simulations 
    is enabled by the parameterization of $\tau_\ssa(t)$ and $\tau_\ffa(t)$ 
    that we give in \S~\ref{sec:tau_param}.
    In \S~\ref{sec:apps} we show how this parameterization can be 
    applied to the planning and interpretation of observations. 
    First, we look at the radio non-detection of PTF11kx, an SN~Ia
    known to interact with a confined, detached shell of CSM.
    We then perform a statistical analysis of radio non-detections of
    SNe~Ia near maximum light to derive the maximum allowed fraction
    of SNe~Ia that can host confined, detached shells.

\section{Summary of Paper I}\label{sec:PaperI}
    
    Paper~I presented a suite of 
    one-dimensional hydrodynamic models of a typical SN~Ia
    interacting with a low-mass, confined shell of CSM.
    This section provides a brief summary of the Paper~I 
    results and  
    reiterates its limitations for the reader's convenience.

    The SN\,Ia ejecta 
    have mass $M_\mathrm{ej} = 1.38~M_\odot$ and 
    energy $E_\mathrm{ej} = 10^{51}~\mathrm{erg}$.
    Before impact with the CSM shell, they are in free expansion.
    This is equivalent to assuming that any CSM within the detached shell is too low density
    to affect the dynamics of the ejecta.
    The ejecta mass-density
    profile is assumed to have a broken power-law structure
    with $\rho_\mathrm{ej} \propto r^{-1}$ in the inner regions
    ($v \lesssim 10,000~\mathrm{km~s^{-1}}$) and 
    $\rho_\mathrm{ej} \propto r^{-10}$ in the outer regions.
    
    The CSM is assumed to be confined to a constant-density shell with 
    density $\rhocsm$ between radii $\Rin$ and $(1+f_R) \Rin$. 
    The parameter $f_R\in[0.1,1]$ is called the ``fractional width'' 
    of the shell, since
    \begin{equation}
        f_R = \Delta R/\Rin ~,
        \label{eqn:fR}
    \end{equation}
    where $\Delta R$ is the width of the shell.

    The ejecta impact the CSM at time $\timp$ after explosion.
    The time of impact is related to $\Rin$ through Paper~I 
    Equation~5,
    \begin{eqnarray}
        \Rin &=& (1.47\times10^{16}~\mathrm{cm})
                 \left( \frac{\timp}{100~\mathrm{days}} \right)^{0.7}        \notag \\
        &&\times                    
               \left( \frac{\rhocsm}{10^{-18}~\mathrm{g\ cm^{-3}}}\right)^{-0.1}
               ~. \label{eqn:Rin}
    \end{eqnarray}
    This scaling ensures that the density ratio between the CSM
    and ejecta at the point and time of first contact is fixed 
    to 0.33, which defines the ``fiducial model set.''

    The models are invalid when (1) $\Rin/\timp > 45,000~\mathrm{km~s^{-1}}$, 
    i.e., they imply an unphysically large ejecta speed,
    (2) $\Rin/\timp < 10,000~\mathrm{km~s^{-1}}$, i.e., interaction is with
    the inner ejecta, or 
    (3) $\rhocsm > 10^{-14}~\mathrm{g~cm^{-3}}$ where cooling and photon trapping 
    are likely to be important, i.e., the adiabatic assumption does not hold.
    At a given $\Rin$, the first constraint places a lower limit on the 
    allowed CSM densities, whereas the second two place upper limits on the 
    CSM density.
    These limitations are summarized in Figure~1 of Paper~I.

    The hydrodynamics are evolved assuming adiabatic evolution
    using the one-dimensional Lagrangian solver of \texttt{SEDONA} 
    \citep{RothKasen15}.
    The hydrodynamic behavior of this system is as follows.
    Initially, the shock ``ramps'' up in the CSM---energy 
    density grows, as does the width of the shock region.
    Before it can reach the self-similar limit, the forward
    shock reaches the edge of the CSM shell---the ``end'' 
    of interaction.
    The hot, accelerated CSM is uncontained by any 
    external material and therefore rapidly expands---a rarefaction wave crosses back toward the ejecta.
    The energy density plummets.

    Paper~I assumes the relativistic electron population
    in the shocked gas is distributed as 
    \begin{equation}
        n_e(E) dE = C_E E^{-p} dE \label{eqn:n_e}
    \end{equation}
    where $E$ is the electron energy, and $p=3$ is assumed. 
    The normalization factor is determined by assuming that the 
    energy density in relativistic electrons is 10\% of the total
    shocked gas energy density, i.e. $\epsilon_e = 0.1$. 
    
    Paper~I shows that the fiducial model set defines a 
    family of optically-thin light-curves.
    The light-curves rise while the shock is in the CSM and
    therefore peak at the time the shock reaches the 
    outer edge of the CSM shell. 
    For this reason, the time the shock reaches the outer
    edge of the CSM is denoted $t_p$. 
    Paper~I Equation~7 gives
    \begin{equation}
        t_p / \timp = 0.983 (1+f_R)^{1.28} ~,
        \label{eqn:x_p}
    \end{equation}
    which can be used to produce an expression for the 
    evolution of the forward shock radius ($R_f$),
    since $R_\mathrm{out}/R_\mathrm{in} = (1+f_R)$, 
    \begin{equation}
        R_f/R_\mathrm{in} = 1.013 (t/\timp)^{0.781} ~.
        \label{eqn:Rf}
    \end{equation}
    The peak luminosity scales like
    \begin{eqnarray}
\notag        \mathcal{L}_{\nu,\mathrm{thin},p} &\propto& \epsilon_e^2 \epsilon_B \nu^{-1} 
                                              \rhocsm^{8/7} \Rin^{3/7}  \\
                                            &&  \times [1-(1+f_R)^{-1.28}] \quad,
    \end{eqnarray}
    as can be seen in Paper~I Equations 11 and 37, 
    where $\epsilon_B$ is the ratio of the magnetic field
    energy density density to the gas energy density and is
    typically assumed to be $\epsilon_B=0.1$ though in this
    work we leave it as a free parameter.

    The shape of the light-curves is described by 
    an asymptotic rise (Paper~I Equation~10) followed by 
    a complex decline.
    The decline is described by the $f_R$-dependent
    time it takes the light-curve to reach characteristic
    fractions of the peak luminosity (Paper~I Table~1 and Equation~12).

\section{Escape Fraction of Photons from a Thin Spherical Shell}
\label{sec:lum_calc}
    In this work we consider the absorption of radio emission from
    synchrotron self-absorption in the emitting region itself
    as well as external absorption by the free-free (Bremsstrahlung)
    process in the external, unshocked CSM.
    The strategy for obtaining optically thick radio light-curves
    from the optically thin light-curves parameterized in Paper~I
    is to simply find the escape fraction of radio photons,
    the ratio of the optically thick to optically thin luminosity.
    
    The expression for the escape fraction depends on the geometry of the 
    emitting and absorbing gasses.
    In our case, the emitting (and self-absorbing) region is a thin, 
    spherical shell. 
    The external, absorbing medium is also a thin, spherical shell 
    and only exists before the forward shock overtakes the edge of the
    shell.

    \citet{Weiler+90} provide an expression for correcting
    optically thin luminosity for internal absorption in the
    emitting medium,
    \begin{equation}
        \mathcal{L} = \Lthin \left(\frac{1-e^{-\tau}}{\tau}\right),
        \label{eqn:slab_approx}
    \end{equation}
    which is the calculation for a planar slab geometry and $\tau$ is the 
    optical depth of the slab along the line of sight.
    In Appendix~\ref{appen:rays}, we show that the full solution for
    a thin shell geometry has the same asymptotic behavior as 
    the slab approximation so long as one uses an appropriate 
    expression for $\tau$. 
    Given the synchrotron self-absorption extinction coefficienct
    ($\alpha_\mathrm{ssa}$) and the volume-to-surface-area ratio 
    of the emitting sphere ($\DeltaR$), the appropriate $\tau$ to 
    capture the effect of synchrotron self-absorption is 
    \begin{equation}
        \tau_\ssa = 4\alpha_\mathrm{ssa}\DeltaR \label{eqn:tau_ssa}
    \end{equation}
    such that the escape fraction in the 
    absence of an external absorbing medium can be approximated by
    \begin{equation}
        \mathcal{L_\nu}/\mathcal{L}_\mathrm{\nu,thin} =  \frac{1-e^{-\tau_\ssa}}{\tau_\ssa} \quad,
    \end{equation}
    as derived in Appendix~\ref{appen:noext}.
    In the presence of an absorbing medium with extinction
    coefficient $\alpha_\ffa$ and 
    radial width $\Delta r_\mathrm{ext}$, we take 
    the free-free optical depth to be
    \begin{equation}
        \tau_\ffa = \alpha_\ffa \Delta r_\mathrm{ext} \label{eqn:tau_ffa}
    \end{equation}
    and approximate the escape fraction as
    \begin{equation}
        \mathcal{L_\nu}/\mathcal{L}_\mathrm{\nu,thin} =  \frac{1-e^{-\tau_\ssa}}{\tau_\ssa} 
        e^{-\tau_\ffa} \quad,
        \label{eqn:esc_frac}
    \end{equation}
    as discussed in Appendix~\ref{appen:ext}.
    As noted in the appendices, the error incurred by using these
    approximations in lieu of the exact integral depends on the 
    extent of the media and their optical depth but is typically
    small.

    Thus, the goal of this work is to determine, from the simulations,
    the time evolution of $\alpha_\mathrm{ssa}\DeltaR$ and 
    $\alpha_\ffa \Delta r_\mathrm{ext}$.

\section{Calculation of Optical Depth from Simulations}\label{sec:tau_calc}

\begin{figure}
        \centering
        \includegraphics[width=\linewidth]{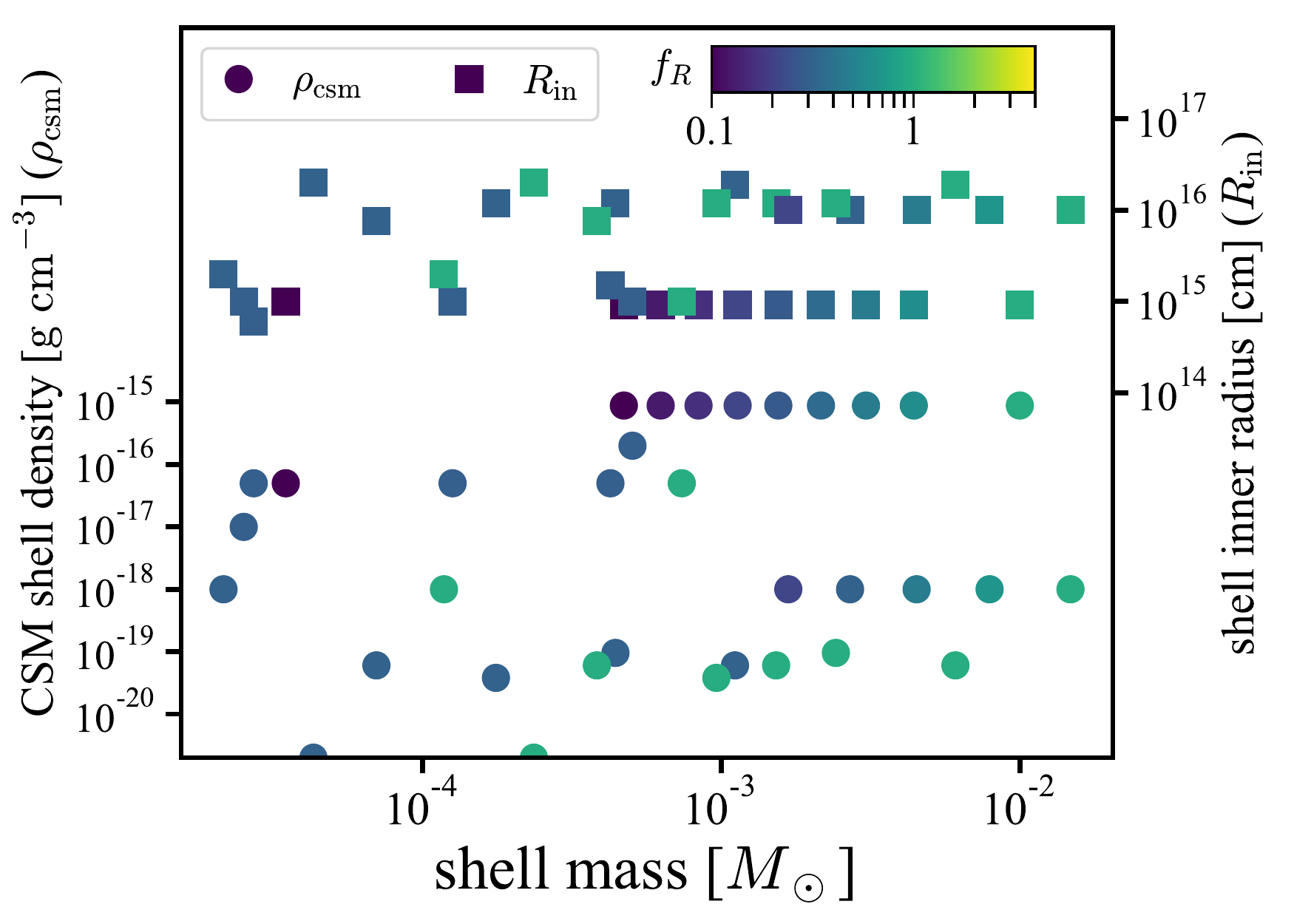}
        \caption{Summary of the CSM shell properties for the simulations
        presented in this work. 
        \textbf{Circles} show the CSM density (left axis) and 
        \textbf{squares} represent the inner radius (right axis), 
        and \textbf{color} illustrates the shell width (color bar). 
        Axis limits are set to separate the density and
        radius points.}
        \label{fig:mod_summary}
    \end{figure}

    In this section we describe how we calculate the optical depths 
    $\tau_\ssa$ and $\tau_\ffa$ from hydrodynamic models.
    Figure~\ref{fig:mod_summary} shows the CSM shell properties
    of the models, which cover a range of shell masses through
    variations in the shell location, extent, and density.
    
    First we must calculate the extinction coefficient of synchrotron self-absorption, $\alpha_{\rm ssa}$, 
    in each shocked resolution element for each simulation
    snapshot.
    We perform these calculations with 
    \texttt{rad\_tools.SynchrotronCalculator} 
    of \texttt{csmpy}.\footnote{\texttt{https://github.com/chelseaharris/csmpy}}

    The synchrotron 
    extinction coefficient ($\alpha_{\rm ssa}$) in each resolution
    element of the simulation is calculated according to the equation in
    \citet{RybickiLightman1979},
    \begin{eqnarray}
\notag    \alpha_{\rm ssa}(\nu) &=&
        \nu^{-(p+4)/2} \frac{\sqrt{3}q_e^3}{8\pi m_e} 
        \left( \frac{3 q_e}{2\pi m_e^3 c^5} \right)^{p/2}\\
        &&\times            
        C_E (2B/\pi)^{(p+2)/2}\\
\notag        &&\times                        
        \Gamma\left(\frac{3p+2}{12}\right)
        \Gamma\left(\frac{3p+22}{12}\right)            ~,
    \end{eqnarray}
    where $p=3$ is the electron distribution power-law index, 
    $q_e$ is the electron charge, $m_e$ the electron 
    mass, $c$ the speed of light, 
    $C_E$ is the normalization of the electron distribution (Equation~\ref{eqn:n_e},
    $B = \sqrt{8\pi \epsilon_B u_{\rm gas}}$ the magnetic field 
    strength 
    (the factor of $2/\pi$ multiplying $B$ in $\alpha_\ssa$ 
    accounts for the pitch angle term as in Paper~I),
    and $\Gamma$ is the gamma function (calculated using
    \texttt{scipy.special.gamma}).

    Equation~\ref{eqn:esc_frac} assumes a constant extinction 
    coefficient in the self-absorbing shell. 
    In reality, especially after the shock crosses the 
    shell and it begins to expand, the extinction 
    coefficient may be different across the shocked gas.
    The representative $\alpha$ we use in our 
    optical depth calculations is the radial average value,
    \begin{equation}
        \langle \alpha_\ssa \rangle = \frac{\sum_k \alpha_{\ssa,k} dr_k}{\sum_k dr_k},
        \label{eqn:al_sim}
    \end{equation}
    where $k$ indicates the index of a resolution element
    in the shock.
    We exclude the five resolution elements closest to the
    contact discontinuity in our calculation of 
    $\langle \alpha_\ssa \rangle$
    because mass-density is a 
    factor in the $\alpha_\ssa$ calculations and is incorrect
    near the contact discontinuity due to the unaddressed
    Rayleigh-Taylor instability, as noted, e.g., in \citet{Chevalier82}. 
    
    The representative shell thickness $\DeltaR$ (Equation~\ref{eqn:dr_repr})
    can be computed directly from the contact discontinuity 
    radius ($r_1$) and the forward shock radius ($r_2$).
    
    Thus for each time snapshot of the simulation we can determine
    \begin{equation}
        \tau_\ssa = 4 \langle \alpha_\mathrm{ssa} \rangle \DeltaR ~.
        \label{eqn:tau_sim}
    \end{equation}

    The free-free extinction coefficient ($\alpha_{\rm ff}$) must
    describe the preshock CSM, which we assume is 
    hydrogen rich, isothermal, constant density, and fully ionized 
    by the radiation field of the shock.
    Therefore, although there are many resolution elements of 
    preshock CSM in the hydrodynamic simulation, 
    for the purposes of radiation transport it is one-zone model.
    The extinction coefficient is calculated using the formulae in
    \citet{RybickiLightman1979} via the  \texttt{rad\_tools.BremCalculator.calc\_al\_BB} 
    function of \texttt{csmpy} which assumes the electrons are
    thermally distributed and uses the gaunt factors calculated by
    \citet{vanHoof+14}. 
    Extinction by the CSM is in the 
    Rayleigh-Jeans limit and the formula used is
    \begin{equation}
        \alpha_\ffa = 0.018 Z^2 g_\mathrm{ff} T_\csm^{-3/2} n_e n_I \nu^{-2}
        \label{eqn:al_ff}
    \end{equation}
    where $Z=1$ is the ion charge, 
    $g_\mathrm{ff}$ is the gaunt factor at the target frequency,
    and $n_e$ and $n_I$ are the 
    electron and ion number densities, which we 
    estimate simply as
    $n_e = n_I = \rho_\csm/m_p$ in this work unless stated otherwise,
    where $m_p$ is the proton mass.
    The width of the pre-shock CSM is 
    \begin{equation}
        \Delta r_\mathrm{ext} = R_\mathrm{out} - R_f \quad,
    \end{equation}
    and thus $\tau_\ffa$ is known from Equation~\ref{eqn:tau_ffa}.

\begin{figure}
        \centering
        \includegraphics[width=\linewidth]{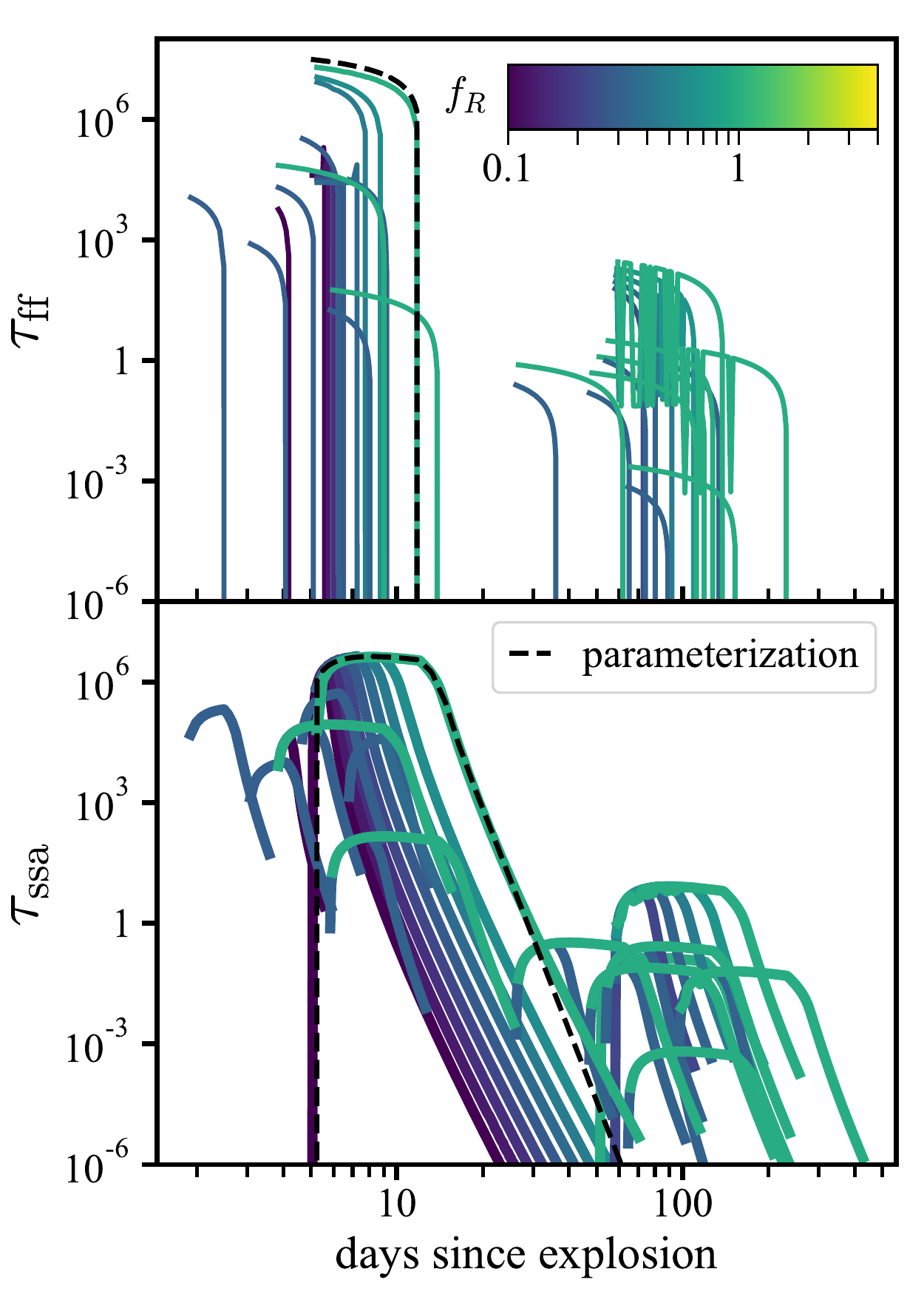}
        \caption{Calculation of $\tau_\ffa$ \textbf{(top)} and 
        $\tau_\ssa$ \textbf{(bottom)} from the models 
        shown in Figure~\ref{fig:mod_summary},
        using $\epsilon_B=0.1$
        (see \S~\ref{sec:tau_calc}).
        Density and impact time simply affect
        the overall position of the evolution 
        (i.e., curves shift up when density increases,
        and to the right when impact time increases), 
        whereas $f_R$  affects the shape of the curve.
        (Note that the jaggedness of some $\tau_\ffa(t)$ curves
        is numerical, not physical)
        The \textbf{black dashed curves} show an example of applying
        the parameterization given in \S~\ref{sec:tau_param}
        for the shell parameters of our highest optical depth model
        ($\rhocsm=8.8\times10^{-16}~\gcc,~f_R=1,~\Rin=9.2\times10^{14}~\mathrm{cm}$).
        }
        \label{fig:tau_nonorm}
    \end{figure}

    The evolution of $\tau_\ssa$ and $\tau_\ffa$ calculated from the
    models are shown in Figure~\ref{fig:tau_nonorm}.
    In these calculations, we have assumed 
    $\epsilon_B=0.1, T_\csm=10^3~\mathrm{K}, \nu=4.9~\mathrm{GHz}, \mu_e=\mu_I=1,$
    and $Z=1$.
    The sharp elbow on the decline of each curve marks $t_p$, 
    the time when the
    forward shock crosses the outer edge of the CSM.
    In the next section, we will discuss what drives the normalization.
    Here we will point out that for most of the time that the shock
    is in the shell, $\tau_\ssa$ and $\tau_\ffa$ are nearly constant. 
    Independent of $f_R$, the models with significant absorption by 
    either process have $\rhocsm\gtrsim10^{-18}~\gcc$. 
    Since $\tau_\ssa(t)$ and $\mathcal{L}_\mathrm{\nu,thin}(t)$
    are driven by the synchrotron process, they have similar shapes, with
    a long tail after the shock has crossed the shell 
    and shells following the same rise independent of $f_R$. 
    Contrary to this, the $\tau_\ffa$ is only important while the 
    shock is in the shell and higher-$f_R$ have higher optical depths at all times.
    In some of the later impact time models, we see that the
    $\tau_\ffa$ curves jump to low values at certain time steps.
    This is a numerical artifact of the shock front identification process
    and does not affect our later results; 
    for the sake of transparency in our methods and because it does 
    not have a large illustrative impact, we have chosen to not to edit
    the $\tau_\ffa(t)$ curves of these models.

\section{Parameterization of Optical Depth Evolution}\label{sec:tau_param}

    In this section, we present a parameterization to allow synchrotron self-absorption
    and free-free (external) absorption optical depths to be reconstructed
    for an arbitrary CSM shell using 
    Equation~\ref{eqn:esc_frac}.
    An example of the light-curves created with this method can be seen in
    Figure~\ref{fig:11kx}, which shows the light-curve with no absorption (dotted curves), 
    $\tau_\ssa$ only (dashed curves), and both sources of absorption (solid curves)
    for models of two different densities.

    The variable for time we will use is 
    \begin{equation}
        x\equiv t/\timp,
    \end{equation}
    i.e., time is normalized to the time of impact.
    The time that the forward shock crosses the edge of the CSM shell is the time
    of peak luminosity in the optically-thin radio light-curves and is denoted by 
    $t_p$ in Paper~I; therefore, here we will use $x_p = t_p/\timp$.
    As in Paper~I, we will provide a functional form for the evolution of 
    $\tau_\ssa$ and $\tau_\ffa$ while $x\leq x_p$, and evaluate times that $\tau_\ssa$
    reaches characteristic values for times $x>x_p$ (at which point there is no
    external absorption because the CSM has been swept over). 

    \subsection{Synchrotron Self-Absorption}
    
    First we determine the normalization of $\tau_\ssa$.
    Using Equation 52 of Paper I, the normalization of 
    the extinction coefficient is
    \begin{equation}
        \alpha_\mathrm{ssa} \propto \rhocsm^{-1} u_\mathrm{gas}^{13/4} \nu^{-7/2} \epsilon_B^{5/4} ~.
    \end{equation}
    From Equation 7 of Paper I, the time evolution of the forward
    shock radius, $R_f$, and shock speed, $v_s$, while the shock is in the shell is
    \begin{eqnarray}
        x &=& 0.983 (R_f/R_{c,0})^{1.28} \\ 
        \Rightarrow
        R_f &=& \Rin x^{0.781} \\
        &\text{and}& \\
        v_s &=& \frac{\Rin}{\timp} x^{-0.219} .
    \end{eqnarray}
    Equation~\ref{eqn:Rin} gives $\Rin$ in terms of 
    $\timp$ and $\rhocsm$.
    Assuming $\ugas \propto \rhocsm v_s^2$, we now have (dropping factors of $x$
    because we are interested in the normalization) 
    \begin{eqnarray}
        \alpha_\mathrm{ssa}(t) 
                              &\propto& \rhocsm^{8/5} \timp^{-1.95} \nu^{-7/2}\epsilon_B^{5/4} ~.
        \label{eqn:al_scale}
    \end{eqnarray}
    The radial term in $\tau_\ssa$ ($\DeltaR$) 
    is the volume-to-area ratio (Equation~\ref{eqn:dr_repr})
    and should roughly evolve like $R_f$ 
    if the shell is thin and has width $\Delta R \propto R_f$.
    Then the expected normalization of $\tau_\ssa$ should scale like 
    \begin{eqnarray}
        \tau_\ssa \propto \alpha_\mathrm{ssa}(t) R_f(t) 
               \propto \rhocsm^{3/2} \timp^{-5/4} \nu^{-7/2} \epsilon_B^{5/4} ~. 
               \label{eqn:tau_scale}
   \end{eqnarray}
   and we find that, indeed, the evolution of $\tau_\ssa(x)$ is the 
   same for all models when normalized by this factor.

   Therefore, $\tau_\ssa(x)$ at times after impact but
   before the shock crosses the outer edge of the CSM (i.e., $1 \leq x \leq x_p$)
   can be described by the asymptotic function
    \begin{eqnarray}
\notag    \tau_\ssa(x) &=& 13.6 \left( \frac{\rhocsm}{10^{-18}~\mathrm{g\ cm^{-3}}}\right)^{3/2}
                    \left( \frac{\timp}{100~\mathrm{days}} \right)^{-5/4} \\
                    &\times&
                    \left( \frac{\nu}{4.9~\mathrm{GHz}} \right)^{-7/2}
                    \left( \frac{\epsilon_B}{0.1} \right)^{5/4}\\
                    &\times&
                    x^{-1.34} (1-x^{-1.66}) \label{eqn:tau_early} ~.
    \end{eqnarray}
    The normalization factor and the exponents in the asymptotic function 
    $x^{-1.34} (1-x^{-1.66})$ were determined using
    \texttt{scipy.optimize.curve\_fit}. 

    As with the optically thin luminosity, we fit the evolution of $\tau_\ssa$ after 
    the shock has crossed the outer edge of the CSM (i.e., $x > x_p$) by determining
    the time at which $\tau_\ssa$ reaches characteristic fractions of 
    $\tau_{\ssa,p} \equiv \tau_\ssa(x_p)$ and connecting the points with 
    power-laws (i.e., linear interpolation in logarithmic space).
    The characteristic points are
    \begin{eqnarray}
        x(\tau_\ssa = 0.5 \tau_{\mathrm{ssa},p}) &=& 1.015 (1+f_R)^{1.38} \label{eqn:tau_mid1}\\
        x(\tau_\ssa = 0.1 \tau_{\mathrm{ssa},p}) &=& 1.046 (1+f_R)^{1.49} \label{eqn:tau_mid2}\\
        x(\tau_\ssa = 10^{-2} \tau_{\mathrm{ssa},p}) &=& 1.118 (1+f_R)^{1.54} \label{eqn:tau_mid3}\\
        x(\tau_\ssa = 10^{-3} \tau_{\mathrm{ssa},p}) &=& 1.206 (1+f_R)^{1.60} \label{eqn:tau_mid4}~. 
    \end{eqnarray}

    Beyond this latest time point we assume adiabatic evolution,
    \begin{eqnarray}
        \alpha_{\rm ssa} &\propto& t^{-3} (t^{-5})^{13/4} \propto t^{-19.25} \\
        \DeltaR &\propto& t \\
        \Rightarrow
        \tau_\ssa &\propto& t^{-18.25} \label{eqn:tau_late} \quad.
    \end{eqnarray}
    where we have assumed the shell inner and outer radii evolve like $r\propto t$, 
    density evolves like $\rho \propto t^{-3}$, and energy density evolves like
    $u \propto V^{-5/3} \propto t^{-5}$, with $V$ being the shell volume.

    \subsection{Free-Free Absorption}
    
    For the external, free-free absorption, 
    the evolution of optical depth reflects the radial evolution of the 
    shock, i.e.,
    \begin{equation}
        \tau_\mathrm{ff}(x) = \alpha_\mathrm{ff} \Delta r_\mathrm{ext}(x) ~.
    \end{equation}
    Using Equation~\ref{eqn:Rf}, this can be estimated as
    \begin{equation}
        \Delta r_\mathrm{ext}(x) = R_\mathrm{in}[(1+f_R) - x^{0.781}] ~,
    \end{equation}
    where we have used $0.987\approx1$ to make it exact at $x=1$ rather than using the fit value, so,
    \begin{equation}
        \tau_\mathrm{ff}(x) = \alpha_\mathrm{ff} R_\mathrm{in}
                             [(1+f_R) - x^{0.781}] ~,
    \end{equation}
    with $\alpha_\mathrm{ff}$ as in Equation~\ref{eqn:al_ff} 
    and $R_\mathrm{in}$ given by Equation~\ref{eqn:Rin}. 
    
    Note that $\tau_\mathrm{ff}(x > x_p) = 0$ because $x_p$ represents 
    the time at which the forward shock crosses the edge of the CSM shell,
    thus, all of the CSM has been shocked and there is no ``external'' medium.

    \subsection{Error of the Parameterization}
    
    The error incurred by using the fitting functions given above
    --- i.e., comparing $\tau_\ssa(x)$ and $\tau_\ffa(x)$ calculated
    with the given formulae versus from the simulations themselves ---
    is small, $\lesssim 30\%$ on each, near the peak of the optically thin
    light-curve.
    Very near the time of impact, when the system is changing rapidly, 
    the error can be much larger.
    We also find that the adiabatic approximation does not match the 
    very late time behavior well, possibly due to deceleration from the
    ``interstellar medium'' gas (of density $10^{-24}~\mathrm{g~cm^{-3}}$)
    that lies outside the shells. 
    It is unlikely that either of these phases will be of practical
    use to the interpretation of observations, since the (optically thin) 
    luminosity of the shocked gas is so low at these times 
    --- $\lesssim0.1\%$ of the optically thin peak luminosity. 
    Nevertheless, we caution that one take care if interpretation of 
    observed data hinges on the very early or late phases of the 
    interaction.

\section{Application to Radio Datasets}\label{sec:apps}

    In this section we show how the parameterized light-curves
    can be applied to radio datasets.
    For these analyses, we assume $T_\csm = 10^4~\mathrm{K}$
    when calculating the free-free absorption, i.e.,
    that the preshock CSM is heated similar to an HII region
    by the ionizing radiation of the shock. 

\subsection{Testing Models of PTF11kx}\label{sec:11kx}
    \citet{Dilday+12} report a non-detection of PTF11kx with the Karl G.\ Jansky Very Large Array (VLA) obtained
    on March 30, 2011 with a $1\sigma$ root-mean-square image noise of $23~\mu$Jy.
    This is +61 days since $B$-band maximum (January 29, 2011). 
    From the NRAO archive, we find that the central frequency of
    the observation was 8.4~GHz.
    
    Consistent with \citet{Graham+17}, 
    in this analysis, we assume a distance of 204.4 Mpc and 
    that $B$-band maximum occurs 13 days after explosion,
    interaction began at $\timp=50~\mathrm{days}$, and 
    interaction ended at $t_p=500~\mathrm{days}$.
    Variations in these timings of $\sim10\%$ do not affect
    our conclusions. 
    Within our model framework, we can derive $f_R$ from $t_p$ and $\timp$
    (Equation~\ref{eqn:x_p}). 
    The only other necessary model input is the density of the shell $\rhocsm$, which can be combined  with $\timp$ to find
    $\Rin$. 

    For PTF11kx, $\rhocsm$ can be estimated from its optical spectra. 
    The Ca~II H\&K absorption lines were saturated
    at early times, allowing an inference of the CSM column density, $N_\csm$,
    assuming solar composition. 
    The mass density, $\rhocsm$, can be found from $N_\csm$ if we assume 
    the density is constant within the shell and if the 
    extent of the CSM ($\Delta R=f_R \Rin$) and mean particle weight ($\bar{m}$)
    are known, as $\rhocsm = \bar{m}N_\csm/\Delta R$.
    We take $\bar{m}$ is 1.33 times the proton mass, as appropriate for neutral material of solar abundance.
    \citet{Graham+17} derive $N_\csm \approx 5\times10^{21}~\mathrm{cm^{-2}}$, 
    significantly lower than the original estimate by \citet{Dilday+12}
    of $N_\csm \approx 10^{23}~\mathrm{cm^{-2}}$.
    The lower estimate is probably correct, for two reasons.
    First, because two different methods for analyzing the line  
    indicate a lower $N_\csm$ \citep{Graham+17}. 
    Second, the higher value of $N_\csm$ creates an inconsistency---to create a saturated line requires that the CSM cover the SN photosphere,
    but full coverage implies a high CSM mass that is inconsistent with the
    weak levels of interaction seen \citep{Dilday+12}. 
    Although we favor the lower density estimate, we will investigate both
    hypotheses.

\begin{figure}
        \centering
        \includegraphics[width=\linewidth]{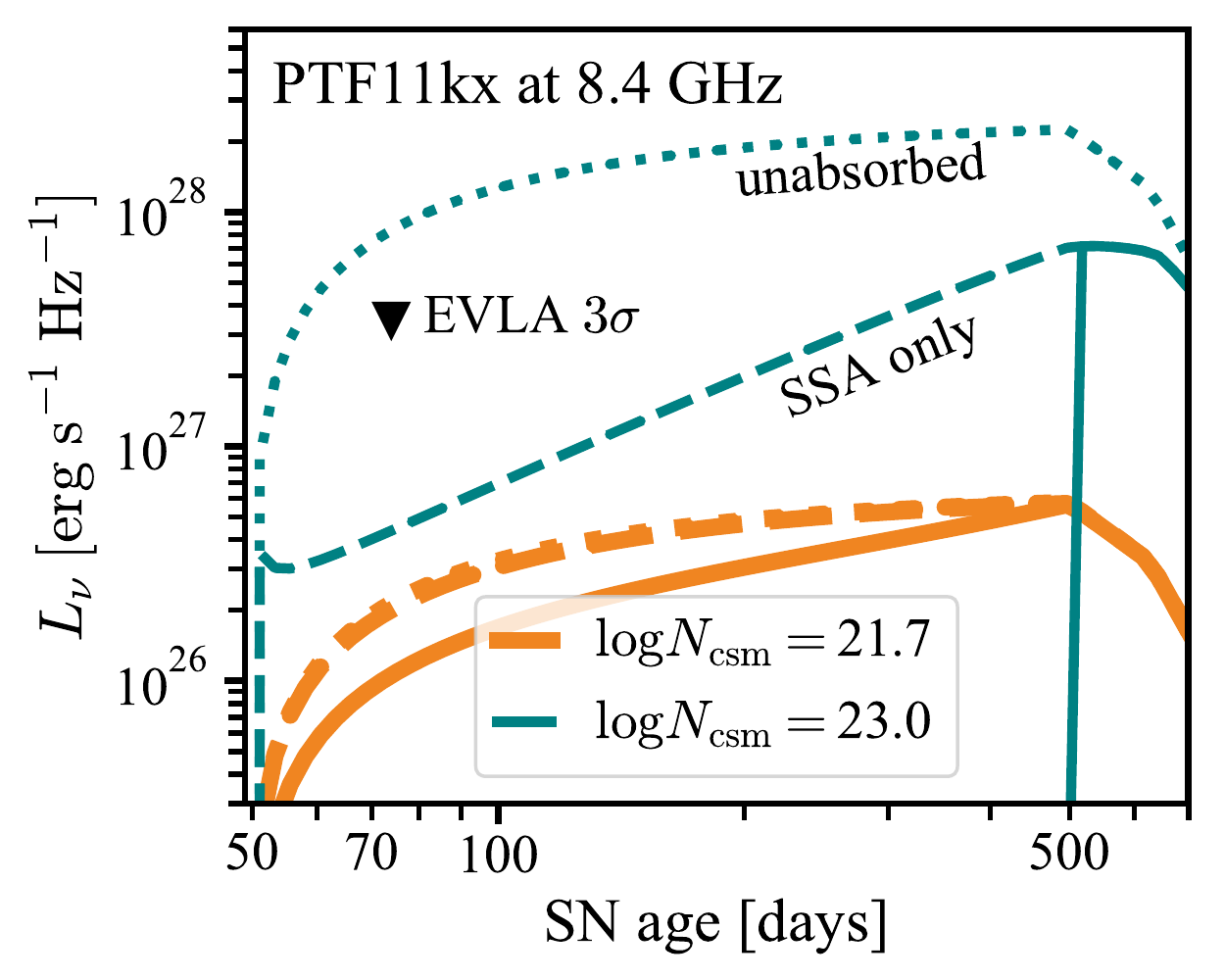}
        \caption{The VLA 3$\sigma$ radio limit for PTF11kx 
        \textbf{(black triangle)} 
        is consistent with models for either 
        $N_\csm \approx 10^{23}~\mathrm{cm^{-2}}$ \textbf{(blue)}
        or 
        $N_\csm \approx 5 \times 10^{21}~\mathrm{cm^{-2}}$        
        \textbf{(orange)}.
        \textbf{Dotted} curves show the optically thin light-curves,
        \textbf{dashed} curves include synchrotron self-absorption ($\epsilon_B=0.1$), 
        and \textbf{solid} curves furthermore include free-free absorption
        (with $Z=1, \mu_e=1.18, \mu_I=1.33, T_\csm=10^4~\mathrm{K}$).
        A radio observation at $\sim500~\mathrm{days}$ may
        have distinguished between the $N_\csm$ measurements.}
        \label{fig:11kx}
    \end{figure}

    In Figure~\ref{fig:11kx} we compare the radio limit for PTF11kx to 
    the radio light-curves based on the current best descriptions of its CSM
    as described above.
    In calculating $\tau_\ffa$, we have assumed $\mu_I=1.33, \mu_e=1.18$ 
    to be consistent with the compositional assumptions used 
    for determining $N_\csm$. 
    We find that with either estimate of $N_\csm$, our models are
    consistent with the radio non-detection of PTF11kx.
    Our optically thick light-curves are needed to interpret the high-$N_\csm$ 
    scenario, whereas the low-$N_\csm$ case is subject to very little absorption.
    If a second observation had been taken around one year after
    explosion, it would have been able to distinguish between the 
    $N_\csm$ values.
    That is, under the assumption of spherically distributed CSM.
    When one allows the CSM to be in a torus, the light-curves must
    be modified; this should roughly be a diminution of the luminosity
    by the covering fraction of the CSM, if we saw PTF11kx edge-on
    as suggested by the saturated pre-impact absorption lines, which would
    make the signal too dim to be seen by the VLA observation.

\subsection{The Allowed Fraction of SNe~Ia with CSM Shells}\label{sec:ensemble}
    Using the optically thick light-curve parameterization (\S\ref{sec:lum_calc}) 
    we can explore the detection power of radio upper-limits for CSM shells. 
    While a similar analysis has been carried out for individual objects 
    \citep{Harris+18, Cendes+20, Pellegrino+20}, this is the first such analysis 
    of a population of SNe Ia.
    In this analysis we assume $\mu_e=\mu_I=1, Z=1,$ and $T_\csm=10^4~\mathrm{K}$.

\begin{figure}
        \centering
        \includegraphics[width=\linewidth]{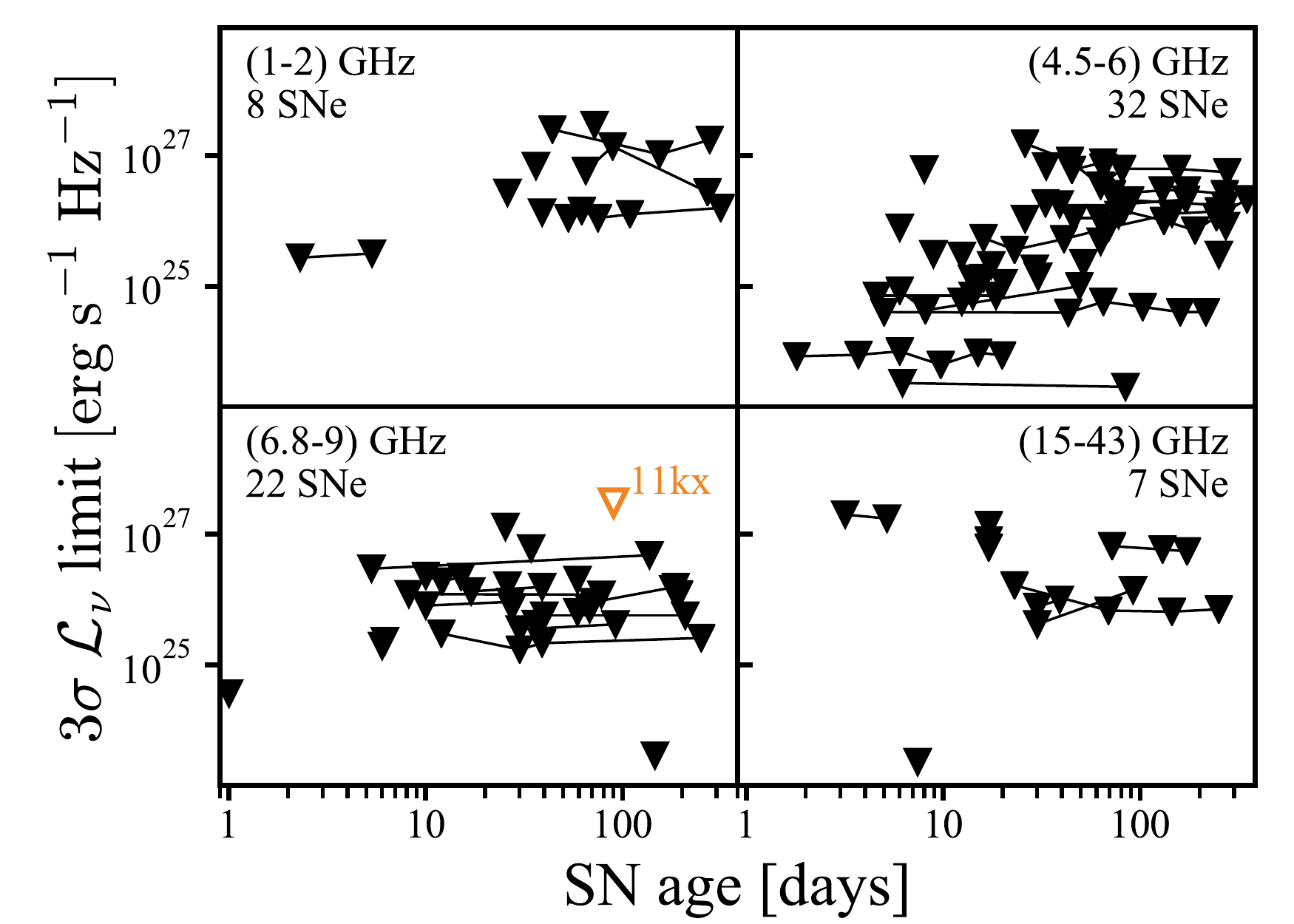}
        \caption{Data ($3\sigma$ upper limits) used to determine the fraction of SNe~Ia
        that may host shells.}
        \label{fig:ensemble_obs}
    \end{figure}

\begin{figure}[t]
        \centering
        \includegraphics[width=.9\linewidth]{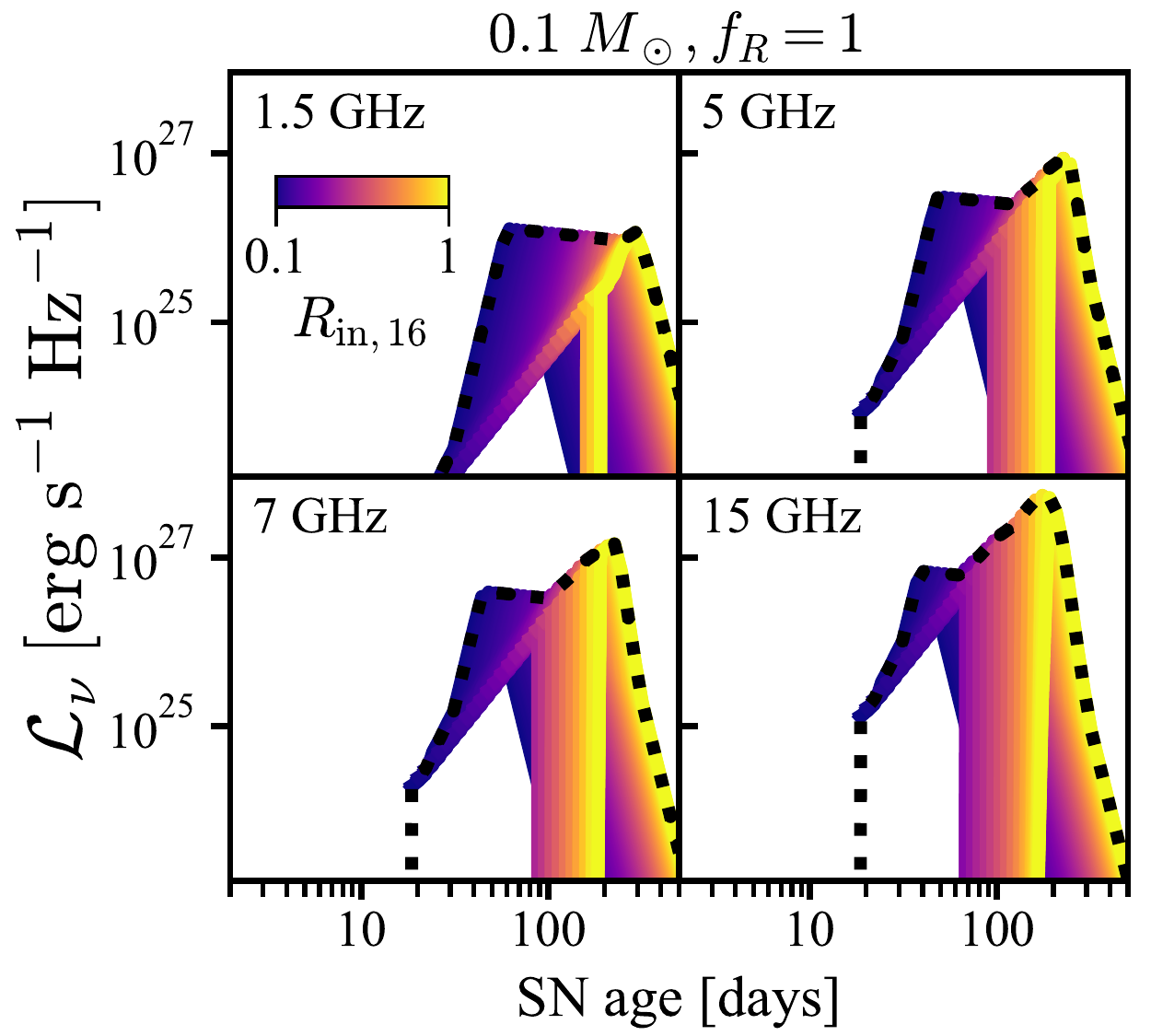}\\
        \includegraphics[width=\linewidth]{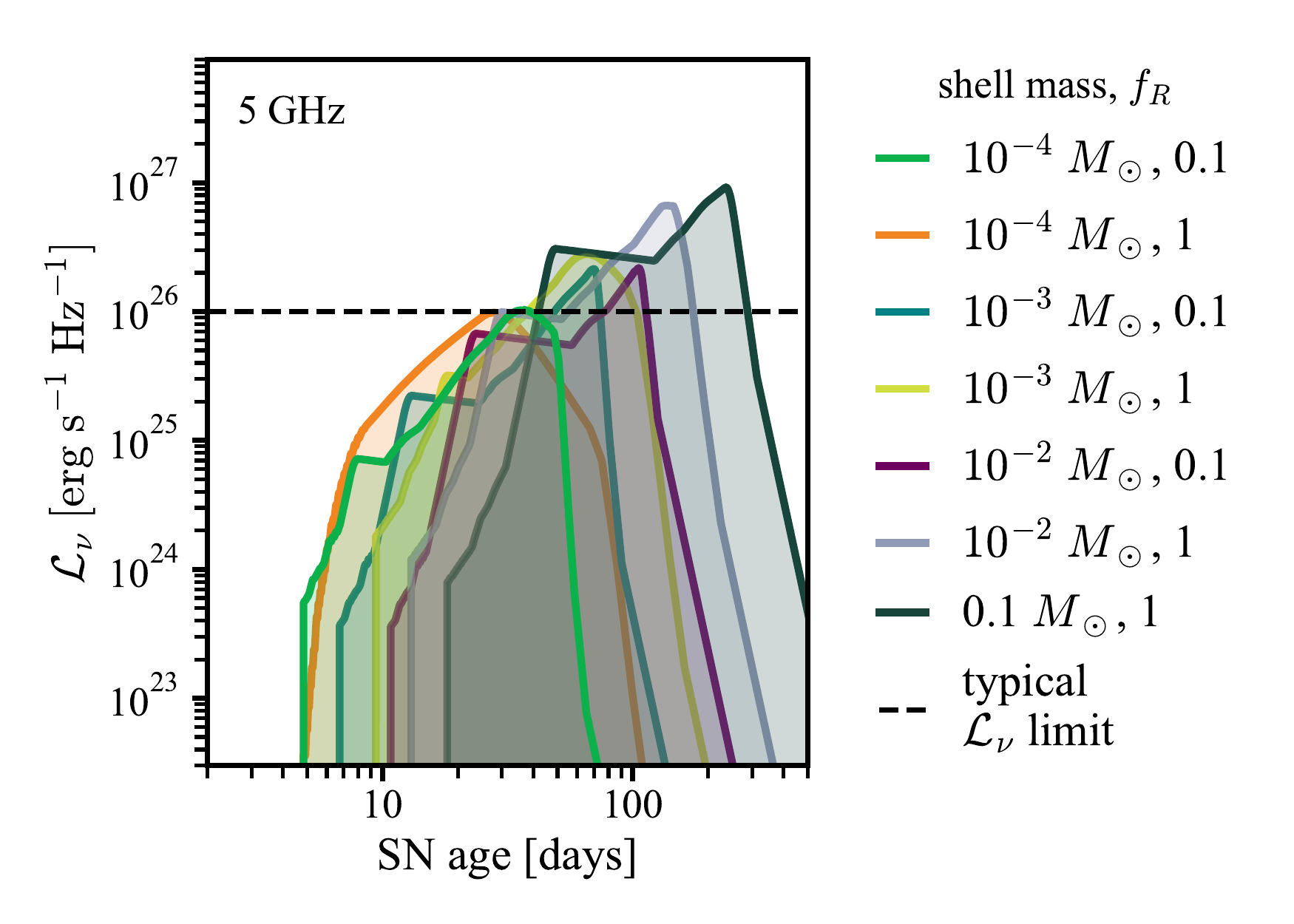}
        \caption{
        \textbf{Top:} A grid of light-curves for a single shell configuration
        ($M_\csm = 0.1~M_\odot$, $f_R=1$), where $R_\mathrm{in,16}$ varies between 0.1--1  (denoted by color scale ranging from magenta to yellow). 
        The light-curve grid forms a ``roof'' at a given frequency \textbf{(black dotted line)}; 
        observations looking to sample this shell configuration must be under the roof.
        \textbf{Bottom:} 5 GHz fiducial light-curves showing roofs for various shell configurations ($R_\mathrm{in,16}\in[0.1,1]$).
        Color represents the shell configuration (see legend). 
        The black dashed line shows the median $3\sigma$ luminosity limit of 
        the SN~Ia sample (Figure~\ref{fig:ensemble_obs}), and the x-axis
        range has been chosen to span the observation times. 
        Observations are primarily under the roofs of $f_R=1$ models, so
        radio observations will be most constraining for these shells.
        }
        \label{fig:shell_envelopes}
    \end{figure}

    \citet{Chomiuk+16} present VLA observations of thermonuclear supernovae
    (SNe\,Ia) across all sub-groups of the class. 
    Of these, we use the ``cool,'' ``shallow-silicon,'' and ``core-normal''
    groups.
    We also incorporate data compiled in \citet{Lundqvist+20} and those
    presented in \citet{Mooley+16} and \citet{Ryder+19}. 
    We group the three sub-types together since the cool and shallow-silicon
    groups are too sparsely sampled to be analyzed independently.
    The sample of data, shown in Figure~\ref{fig:ensemble_obs}, covers a range of frequencies from 1--43 GHz. 
    Observations span $1-365$ days after explosion for a total of 50 SNe
    among all observations independent of frequency (observations are grouped
    by frequency here for comparison with Figure~\ref{fig:shell_envelopes}). 
    We show PTF11kx in this figure for reference; 
    this data point is not included in our following
    statistical analysis because it is not a useful limit, 
    i.e., it would not be able to detect any model in our set of interest.

        In this analysis, we characterize CSM shells by their mass ($M_\csm$), 
        inner radius ($\Rin$), and
        fractional width ($f_R\equiv \Delta R/\Rin$).
        These three parameters fully determine a shell light-curve in our
        model framework.
        We are interested in constraining the fraction of SNe~Ia with CSM 
        of a given $M_\csm$ and $f_R$, which we will call the CSM's ``configuration.'' 
        
        We must choose a distribution of $\Rin$ to make this constraint, which
        we do as follows. 
        Note that we will use $R_\mathrm{in,16} = \Rin/(10^{16}~\mathrm{cm})$.
        \citet{MooreBildsten12} used analytic calculations to explore the CSM established by recurrent nova eruptions in a binary system with a red giant companion and significant associated winds.
        They found that the nova ejecta sweep up the giant wind and 
        quickly (within 20 years) and decelerate 
        to a drastically reduced coasting speed of $\lesssim100~\mathrm{km~s^{-1}}$. 
        The exact values depend on the recurrence time and companion wind mass-loss rate.
        Due to the low speed, the shells build up into a thicker, more massive
        shell than would be formed from an individual nova eruption.
        This slow shell is formed at a distance  
        $R_\mathrm{in,16}\sim0.1-10$, depending on the binary parameters.
        Traveling at $\lesssim100~\mathrm{km~s^{-1}}$, the thick shell will 
        remain in the system for $10^4-10^5~\mathrm{yr}$ before mixing into
        the interstellar medium --- giving plenty of time for a massive shell to
        build up if the recurrent novae continue.

        We consider the delay time between shell formation and SN explosion
        to be entirely unknown (i.e., that the SN event is equally likely to
        occur at any time after the start of the recurrent nova period begins). 
        Therefore, the probability distribution for $\Rin$ is 
        determined by the shell kinematics.
        Since the shells spend only 20 years within $r\sim10^{15}~\mathrm{cm}$
        compared to the $>10^4$ years they spend beyond this distance, 
        we treat the probability of $R_\mathrm{in,16} < 0.1$ as zero.
        Because a shell coasts at constant speed, all radii
        $R_\mathrm{in,16} \geq 0.1$ are equally likely. 
        We only analyze the probability of SNe~Ia having shells with
        $R_\mathrm{in,16} \in [0.1,1]$, i.e., the range that can be 
        studied through observations within a year of explosion 
        (the data; \citealt{Chomiuk+16} specifically limited their survey to 
        radio observations obtained in the first year following explosion). 
        Note that because the shells are freely expanding, $\timp \propto \Rin$
        and $f_R (= \Delta R / R \propto t / t)$ is constant as it moves away from the binary. 
        
        The radio light-curves for the shell configurations we consider 
        are summarized in Figure~\ref{fig:shell_envelopes}.
        
        In the top panel, we show all light-curves generated for
        just one shell configuration in different frequency bins.
        In the actual analysis, a light-curve is generated
        at the the frequency of each individual observation.
        To determine if an observation has constraining power in a
        situation where the location of the CSM is unknown, one must
        not compare a luminosity limit to a single light-curve but instead
        look at the light-curve ``roof'' that is created by the set of
        possibilities (black dashed line).
        Observations under the roof have constraining power, whereas anything
        above the roof has no possibility of detecting any shell and
        therefore no statistical power.
        
        In the bottom panel, we show the roofs for other shell 
        configurations, spanning $M_\csm = 10^{-4}-0.1 M_\odot$ and $f_R = 0.1-1$. 
        The dark green curve (farthest right)
        is the same as the black dotted line from the top panel at 5 GHz, 
        the frequency with the most SNe observed.
        Observations in the shaded regions (i.e., under the roof) 
        have constraining power on the configuration.
        The black dashed line shows the typical luminosity limit of 
        the data, which is very close to
        the top of the ``roof'' for most models, and has best coverage for the
        higher-mass $f_R=1$ shells -- i.e., these are the shells most suited for
        study by the radio observations. 
        Again we note that the PTF11kx observation shown in 
        Figures~\ref{fig:11kx} and \ref{fig:ensemble_obs} is 
        above the roofs, so it is not useful to our analysis.
        Note that CSM parameters that would violate model assumptions 
        (as described in \S~\ref{sec:PaperI}) are not shown, 
        which is one reason that the low-mass shell roofs look different from 
        those of higher masses. 

        The probability of detecting interaction with a CSM shell of mass $M_\csm$ and 
        fractional width $f_R$ in an SN~Ia event is the product of
        (a) the fraction of SNe~Ia that host such shells ($\xi$) and
        (b) the probability that observations of the SN can detect the interaction signal ($\mathcal{P}_i(\text{det}|\text{occ})$).
        The former term is the one of interest to our study, and it can range from $\xi \in [0,1]$.
        
        The latter term we calculate by creating a grid of model light-curves with
        $R_\mathrm{in,16}\in[0.1,1]$, then injecting these
        light-curves into the set of $3\sigma$ upper-limits for the observed SN sample (i.e., Figure \ref{fig:ensemble_obs}).
        The number of injected light-curves that would be detectable for SN $i$ ($N_{\mathrm{det},i}$)
        compared to the number in the grid ($N$) is a good estimation of the detection 
        probability for the shell, provided that the number of models is high enough: 
        \begin{equation}
            \mathcal{P}_i(\text{det}|\text{occ}) = \frac{N_{\mathrm{det},i}}{N}
        \end{equation}
        We use $N=100$ models.
        Thus for each SN the probability of detecting interaction is
        \begin{equation}
            \mathcal{P}_i(\text{det}) = \xi \frac{N_{\mathrm{det},i}}{N} \quad .
        \end{equation}

        The probability that of $S$ events, \textit{none} discovered interaction with
        a shell is
        \begin{equation}
            \mathcal{P}(\text{no dets}) = \prod_{i=0}^{S} \left( 1 - \xi \frac{N_{\mathrm{det},i}}{N} \right) ~.
        \end{equation}
        
        In a Bayesian framework, the probability density $p$ of a given value of $\xi$ being
        true is
        \begin{equation}
            p(\xi | \text{no dets}, I) \propto p(\text{no dets}|\xi, I) \times p(\xi | I),
        \end{equation}
        where $I$ represents our model assumptions.
        Since our model assumptions do not depend on the fraction of SNe~Ia with CSM shells,
        $p(\xi|I) = 1$. 
        The probability density $p(\text{no dets}|\xi, I)$ is proportional to
        $\mathcal{P}(\text{no dets})$. 

        Thus the observed non-detections can be transformed into an upper limit on $\xi$ via
        \begin{equation}\label{eq:xi_up}
            \mathcal{P}(\xi < \xi_\mathrm{up}) = \frac{ \int_0^{\xi_\mathrm{up}} \prod_{i=0}^{S} \left( 1 - \xi^\prime \frac{N_{\mathrm{det},i}}{N} \right) d\xi^\prime}{\int_0^1 \prod_{i=0}^{S} \left( 1 - \xi^\prime \frac{N_{\mathrm{det},i}}{N} \right) d\xi^\prime}
        \end{equation}
        and we can obtain the maximum allowed value of $\xi$ at  99.7\%-confidence ($3\sigma$)  by finding the $\xi_\mathrm{up}$ at which Equation \ref{eq:xi_up} evaluates to 0.997. 
        
\begin{figure}
        \centering
        \includegraphics[width=\linewidth]{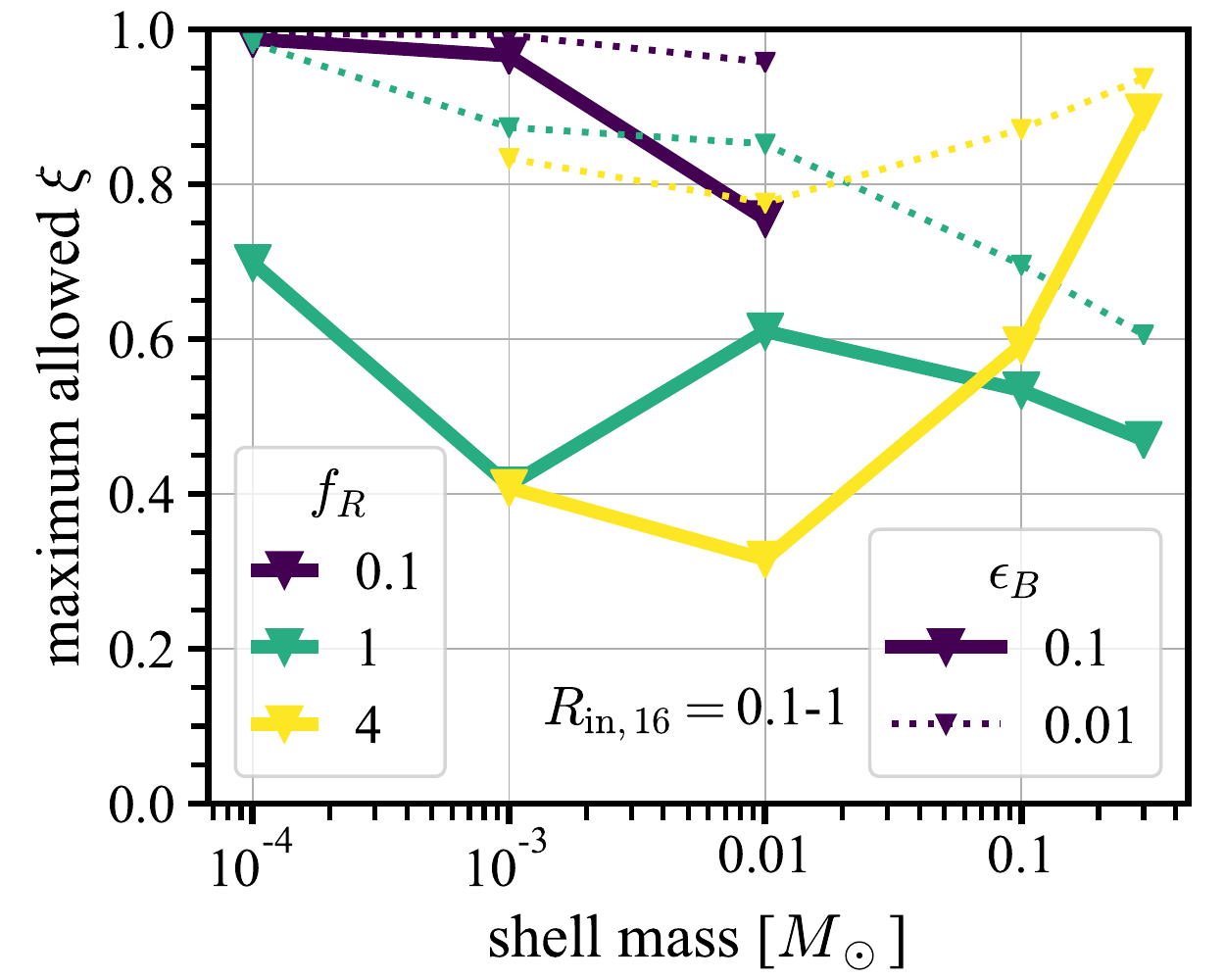}
        \caption{Maximum fraction of SNe~Ia ($\xi$) that can have CSM shells of a given mass and width, within $R_{\rm in} = 10^{16} {\rm cm}$.
        The value of $\xi$ is calculated at 3$\sigma$  (99.7\%) confidence.
        \textbf{Large markers} are calculations with $\epsilon_B=0.1$, while
        \textbf{small markers} represent $\epsilon_B=0.01$. 
        The shell with $M_\csm=0.3~M_\odot$ and $f_R=4$ represents a PTF11kx-like
        configuration.
        }
        \label{fig:Ia_frac_limits}
    \end{figure}

        Figure~\ref{fig:Ia_frac_limits} shows the results of the analysis, 
        providing 99.7\% confidence limits on $\xi$, assuming 
        $\epsilon_B = 0.1$ (large markers, solid lines) and 
        $\epsilon_B = 0.01$ (small markers, dotted lines).
        We see that in all cases, the large sample of radio non-detections is
        still consistent with a high fraction of SNe~Ia having confined CSM
        shells --- especially if $\epsilon_B=0.01$ is the appropriate value for
        these shocks.

        \paragraph{Single nova outbursts.}
        The thickness of a single nova outburst is predicted to be $f_R\sim0.1$ in
        \citet{MooreBildsten12}; these should also be low mass 
        \citep[e.g., $\sim10^{-4}~M_\odot$][]{Chomiuk+14}.
        We see that nova-like shells (thin and low-mass) are currently
        largely unconstrained by radio observations.
        Essentially all SNe~Ia could have a $10^{-4}~M_\odot$, $f_R=0.1$ shell 
        hiding in their circumstellar environment, according to these radio data
        as interpreted in our model framework.
        Nova ejecta spread over a larger volume ($f_R=1$) are only constrained
        to $\lesssim70\%$ of all SNe~Ia, assuming $\epsilon_B=0.1$. 
        
        \paragraph{Multiple novae.}     
        Multiple nova eruptions could produce a thicker, more massive shell. 
        We see that (for $\epsilon_B=0.1$), $f_R=1$ shells are constrained to be
        $\lesssim 50\%$ across the mass range explored.
        This is because, as can be seen in Figure~\ref{fig:shell_envelopes}, 
        these models have similar requirements for their observability ---
        higher mass shells are more luminous (in the optically thin limit) but 
        are also subject to more absorption. 

        \paragraph{Very thick (PTF11kx-like) shells.}        
        In our analysis, we include a $0.3~M_\odot$ shell with $f_R=4$, 
        representing a PTF11kx-like CSM \citep{Graham+17}.
        From the radio limits alone, we find that up to 90\% of SNe~Ia 
        could have CSM with a PTF11kx-like configuration. 
        Lower mass, thick shells are more constrained because they have a lower
        free-free optical depth, yet we see that the radio non-detections 
        are still consistent with a relatively high fraction of SNe~Ia
        having thick shells.
        
        \subsubsection{Comparison with Nebular H$\alpha$ Statistics}

        Recently, \citet{Tucker+20} used a sample of 111 low-redshift 
        SNe~Ia to constrain
        the dominance of the single-degenerate channel in creating SNe~Ia
        using the theoretical framework of \citet{Botyanszki+18}.
        These models focus on the H$\alpha$ signature from hydrogen that
        has been stripped off the companion envelope, and they allow 
        one to convert flux limits into limits on the mass of stripped material
        (subject, of course, to a variety of underlying model assumptions), 
        which can then be compared to theoretical expectations.
        
        One striking decision made in the \citet{Tucker+20} analysis was 
        to exclude all known cases of SNe~Ia with late-time H$\alpha$ emission, even 
        if those events looked normal near maximum light and had observations 
        in the same time frame as the rest of the sample --- 
        the 91T-like (``shallow silicon'') PTF~11kx and the 91bg-like (``cool'') 
        events SN~2018fhw and SN~2018cqj \citep{Dilday+12,Kollmeier+19,Prieto+20}.
        Another example of a 91T-like SN~Ia with late-time H$\alpha$ emission, 
        but that unfortunately does not have observations in the 
        time window considered, is SN~2015cp \citep{Graham+19}.
        \citet{Tucker+20} essentially argue that these events ought to be excluded because
        they are abnormal; but, by definition, hydrogen emission is
        abnormal in any SN~I. 
        We note that there is a strong distinction between the 91T-like delayed-interaction
        events and 91bg-like cases; pertinent to this discussion,
        91bg-like cases have low-luminosity line emission and 
        may represent stripped companion material,
        whereas 91T-like cases have higher line luminosity and a distinct CSM origin.
        Therefore, it was sensible for \citet{Tucker+20} not to analyze PTF11kx or SN~2015cp 
        in the stripped-companion model framework, since the line signal was not of this origin.
        However, we note that the 91bg-like events had estimated stripped masses of 
        $\sim10^{-2}-10^{-3}~M_\odot$, and we estimate that the \citet{Tucker+20} have 65 events 
        that probe a similar mass (their Figure~6), therefore had these events been 
        included in the nebular sample the statistics would have been two detections
        among 67 events, which under a simple binomial distribution analysis results
        in a $3\sigma$ limit of $(0.3-13.4)\%$ of SNe~Ia with $\sim10^{-2}-10^{-3}~M_\odot$
        of (stripped) hydrogen.
        
        Our study, however, is not concerned with the signature of stripped material, 
        but rather with circumstellar material. 
        We do not yet have H$\alpha$ emission models for the CSM shell 
        scenario investigated in this work, so we cannot perform an analysis 
        we have done for the radio using the \citet{Tucker+20} data. 
        However, what we can say is that any of the \citet{Tucker+20} observations
        would have been able to detect H$\alpha$ emission from a PTF11kx twin.
        Including the other three events into the statistics is complicated by their
        lower luminosity (SN~2018cqj), observations being earlier than the rest of the
        sample (SN~2018fhw), or observations being later than the rest of the sample
        (SN~2015cp). 
        If \citet{Tucker+20} had chosen to include PTF11kx in their sample, 
        then they would have 104 normal, 91T-like, or 91bg-like events in their sample 
        (the categories we analyze in this study)
        and one detection of H$\alpha$ emission in the 3--15 months after maximum light
        time window, resulting in an allowed fraction of of SNe~Ia with CSM like PTF11kx 
        of $(0.03-7.21)\%$ at 99.7\% confidence.
        In comparison, our limit from radio data is that up to $\sim90\%$
        of SNe~Ia could have a PTF11kx-like shell (because, for the majority of the interaction,
        we predict the radio emission is absorbed by the preshock CSM). 

        One point of interest to both the radio and optical studies of delayed interaction
        is that hydrogen emission in SNe~Ia --- regardless of its time of appearance --- 
        is so far associated with 91bg-like or 91T-like SNe~Ia 
        \citep[see the above references for individual events with late-time hydrogen emission as well as][]{Leloudas+15},
        which are relatively rare.
        Therefore, even large samples, like those discussed in this work, will 
        not provide a statistically significant number of events from these subgroups. 
        For example, if PTF11kx had been included in \citet{Tucker+20}, 
        the sample size of 91T-like events would be six, with one detection, and
        the prevalence of PTF~11kx-like objects constrained to (0.8--77.1)\% of 91T-like SNe~Ia ---
        or (0.6-79.9)\%, if SN~2015cp is included also.
        For 91bg-like events, the sample would become ten events with two detections,
        and the fractional limit constrained to (0.9-75.6)\% of 91bg-like events similar to 
        SN~2018fhw.
        If these subgroups represent the single-degenerate channel, as has been suggested 
        on theoretical grounds by \citet{FisherJumper15}, 
        then these statistics highlight how little we know about SNe~Ia that do come from 
        the single-degenerate channel compared to the constraints that have been made on
        the prevalence of the single-degenerate channel overall. 
        Furthermore, no SN~Ia with hydrogen emission fits neatly into the single-degenerate
        progenitor picture (having either too much or too little hydrogen mass inferred, and 
        nothing that looks like a normal stellar wind) which challenges our picture of 
        this pathway to explosion --- and therefore, challenges some of the very models used to
        constrain its prevalence among SNe~Ia.

\section{Summary}\label{sec:summary}

SNe~Ia with detached, confined shells of CSM (which produce SNe~Ia;n) 
provide a window into the single-degenerate channel and may represent
SNe~Ia impacting a CSM shaped by novae.
However, the uncertain mass, extent, and location of these shells
makes it challenging to observe them in an interacting phase, creating
large uncertainty in the intrinsic prevalence of these shells.
Adding to this uncertainty, and what this work aims to alleviate, 
is the need for theoretical tools that can interpret SN~Ia observations 
in the context of interaction with these shells --- 
because observations are taken during periods of hydrodynamic transition, 
popular equations based on asymptotic solutions cannot be accurately applied
and new ones must be found.
Without appropriate modeling, the properties of these shells cannot be
precisely determined (limiting studies of their origin), nor can their
occurrence rate be assessed from an SN~Ia survey.

In \citet[][Paper~I, summarized in \S~\ref{sec:PaperI}]{HNK16}, 
we presented hydrodynamic models of shell interaction scenarios
for thin, low mass shells and their corresponding optically thin
synchrotron radio light-curves.
We found a parameterization to reproduce the light-curve of a shell
interaction given the shell properties.
In this paper, we have extended those results to account for
synchrotron self-absorption and free-free absorption.
This allows an exploration of higher-density shells than was possible 
from the results of that work. 

In \S~\ref{sec:lum_calc} we describe our method for using the optical
depth to synchrotron self-absorption ($\tau_\ssa$) and 
free-free absorption ($\tau_\ffa$) to obtain the escape fraction of
the radio photons. 
We find that around a shell density of $\rhocsm\gtrsim10^{-18}~\gcc$,
both sources of absorption begin to come into play. 
We then derive 4.9~GHz $\tau_\ssa$ and $\tau_\ffa$ values from the 
hydrodynamic model suite, 
which requires finding the shock width and mean extinction coefficient
as a function of time (\S~\ref{sec:tau_calc}), to 
explore how these quantities evolve over time.
In \S~\ref{sec:tau_param} we showed how the optical depth evolution
calculated from the simulations
can be parameterized in a similar way to the optically thin light-curves,
allowing for these quantities to be calculated once the shell
properties are specified.
For convenience of use, these 
parameterizations are implemented in a Python script 
(\texttt{HNK16\_tools.py}) available online.
\footnote{\texttt{https://github.com/chelseaharris/csmpy}}

In \S\ref{sec:apps} we apply this new tool to radio observations of 
SNe~Ia. 
First, we consider the radio non-detection of PTF11kx \citep{Dilday+12} 
and assess whether it is consistent with the current picture of 
its CSM \citep{Silverman+13,Graham+17} --- a $\sim 0.3~M_\odot$ 
shell extending from
$\sim10^{16}~\mathrm{cm}$ to $\sim5\times10^{16}~\mathrm{cm}$. 
We find that the radio non-detection is consistent with this model, 
and the non-detection limit was well above the maximum radio
luminosity reached by the interaction at any phase.
\citet{Dilday+12} originally proposed a higher density of CSM, 
and we show that (if this had been spherically distributed) 
it would have reached a detectable level,
but only at late times ($\sim1~\mathrm{year}$ post-explosion), 
when free-free absorption no longer played a role,
which may be a worthy consideration for future radio studies of 
SNe~Ia with stronger interaction.

Second, we use the optically thick light-curve models to
statistically assess an ensemble of SN~Ia radio non-detections
at various times and frequencies.
The parameter of interest is $\xi$, the fraction of SNe~Ia that host
a shell of mass $M_\csm$ and fractional width $f_R$ 
at a distance of $10^{15}-10^{16}~\mathrm{cm}$.
We consider $M_\csm$ between $10^{-4}~M_\odot$ and
$0.3~M_\odot$ with $f_R=0.1,1,$ and $4$. 
Overall, we find that, at 99.97\% statistical confidence, 
thick shells ($f_R=1, 4$) of any mass $<0.1~M_\odot$ 
can be present in up to $\xi\sim60\%$ of SNe~Ia and still be consistent 
with the radio non-detections.
Thin shells are essentially completely unconstrained.
Surprisingly, PTF11kx-like shells, which should be relatively easy to 
see in optical spectra, are only constrained by radio data to 
be in $\lesssim90\%$ of SN~Ia systems, because these relatively massive and 
thick shells are more subject to free-free absorption. 
We further calculate the constraints under the assumption of weaker
magnetic field amplification $\epsilon_B=0.01$, in which case the
radio limits allow a large majority of SNe~Ia to host shells.

\acknowledgments

C.E.H.\ and L.C.\ are grateful for support from NSF through AST-1751874 and AST-1907790. C.E.H.\ also acknowledges support from the Packard Foundation.P.E.N. acknowledges support from the DOE under grant DE-AC02-05CH11231, Analytical Modeling for Extreme-Scale Computing Environments.

\added{We thank the anonymous reviewer for their comments on this manuscript.}

This research used resources of the National Energy Research Scientific Computing Center, a DOE Office of Science User Facility supported by the Office of Science of the U.S. Department of Energy under Contract No. DE-AC02-05CH11231.

The National Radio Astronomy Observatory is a facility of the National Science Foundation operated under cooperative agreement by Associated Universities, Inc.

Michigan State University occupies the ancestral, traditional, and contemporary Lands of the 
Anishinaabeg–Three Fires Confederacy of Ojibwe, Odawa, and Potawatomi peoples. 
The University resides on Land ceded in the 1819 Treaty of Saginaw.

%

\vspace{5mm}
\facilities{Karl G.\ Jansky Very Large Array}


\software{
Sedona \citep{Kasen+06},
SciPy \citep{SciPy}, 
NumpPy \citep{NumPy}, 
Astropy \citep{Astropy},
Matplotlib \citep{Matplotlib}
}



\appendix

\section{Ray Tracing in a Spherical Shell Geometry}\label{appen:rays}
    In the optically thin limit of isotropic emission,
    the spectral (or ``specific'') luminosity of an emitting shell can be
    simply calculated as 
    \begin{equation}
        \mathcal{L}_{\nu,\mathrm{thin}} = 4\pi j_\nu V,
    \end{equation}
    where $j_\nu$ is the emissivity (units like $\mathrm{erg~s^{-1}~Hz^{-1}~cm^{-3}~str^{-1}}$, 
    value is assumed constant throughout the shell), 
    $V$ is the volume (units like $\mathrm{cm^{3}}$), and the factor of $4\pi$ accounts for the angle
    covered by the emission (thus has units of str [sterradians]).
    In the case that the shell both emits and absorbs light, however, one must solve the 
    radiation transport equation.

    \subsection{Internal Absorption Only}\label{appen:noext}

    Consider a thin, spherical shell extending from radius $r_1$ to $r_2$, with constant extinction coefficient 
    ($\alpha$, units like $\mathrm{cm^{-1}}$) and $j_\nu$ 
    within the shell and no emission or absorption in the cavity $r<r_1$.
    This is the scenario if one assumes (1) absorption by the unshocked ejecta is negligible and
    (2) absorption by the unshocked CSM is negligible, either due to the CSM column density or 
    because the shock has already crossed the outer edge of the CSM shell.
    (Note that even if the ejecta full absorb the radio emission, it will
    have a negligible effect on the overall radio luminosity because most 
    of the emission comes from the projected inner edge of the radiating shell,
    as is familiar from spatially-resolved examples of interaction, such as
    the H$\alpha$ emission of SN remnants or radio 
    interferometry of interacting SNe.)
    
    \begin{figure}
        \centering
        \includegraphics[width=\linewidth]{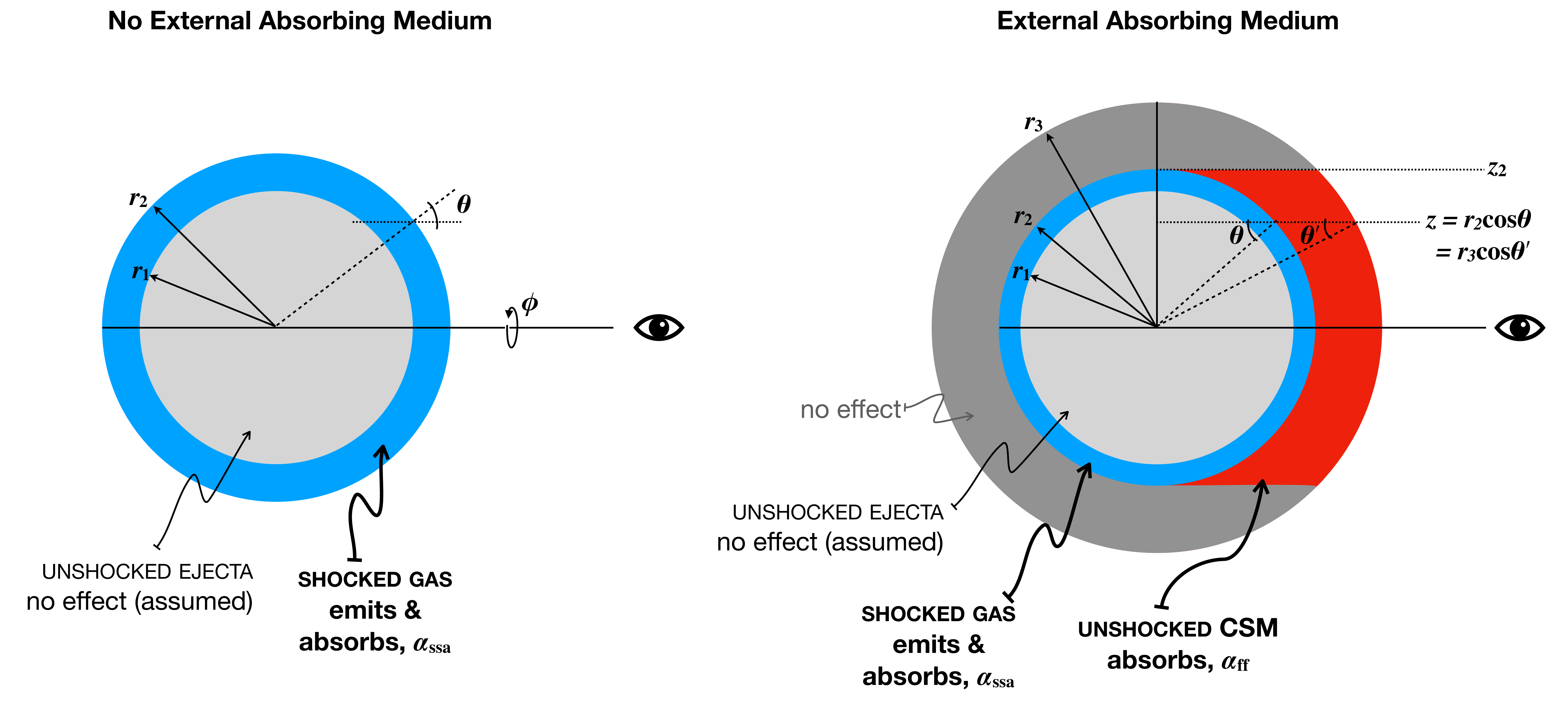}
        \caption{Schematic of the geometries considered for solving the
        radiation transport equation: the case with no external absorption
        (\textbf{left}) and with external absorption (\textbf{right}).
        The ejecta absorption is assumed to have no effect
        because the contribution of the obscured emission region to the
        overall luminosity is small.
        }
        \label{fig:RTdiagram}
    \end{figure}

    Then the solution to the radiation transport equation along a straight path through the 
    sphere that makes an angle $\theta$ with the outward surface normal at $r_2$ is
    \begin{equation}
        I_\nu(\theta) = \frac{j_\nu}{\alpha} \{1-\exp[-\tau(\theta)]\} 
        = \frac{L_{\nu,\mathrm{thin}}}{4\pi V \alpha} \{1-\exp[-\tau(\theta)]\} ~,
        \label{eqn:I_Sconst}
    \end{equation}
    where $I_\nu$ is the specific intensity (units like $\mathrm{erg~s^{-1}~Hz^{-1}~cm^{-2}~str^{-1}}$),
    $\tau$ is the optical depth ($\alpha$ multipied by the path length),
    and in the second expression we have substituted in the equation for optically thin luminosity.
    Defining $\tau_2 \equiv \alpha r_2$ and  
    $\theta_1$ by $\sin\theta_1 \equiv r_1/r_2$, and using the 
    convention $\mu \equiv \cos\theta$ (thus $\mu_1 = \cos\theta_1$), 
    the optical depth is given by
    \begin{eqnarray}
        \tau(\theta) = \left\{ 
        \begin{array}{lr}
        2\tau_2 \mu &  \quad \text{if}\ \mu \leq \mu_1  \\
        2\tau_2 (\mu - \sqrt{\mu^2 - \mu_1^2}) & \quad \mu > \mu_1 
        \end{array} 
        \right.                       
        \quad .
        \label{eqn:tau_shell}
    \end{eqnarray}

    The emerging luminosity from the surface of the sphere ($r=r_2$) is 
    \begin{equation}
        \mathcal{L}_\nu = 4\pi r_2^2 F_\nu = 4\pi r_2^2 \oint I_\nu(\mu)\, \mu\, d\mu\, d\phi ,
    \end{equation}
    where $\phi$ is the angle in the plane perpendicular to the line of sight.
    Since we are considering isotropic, spherical emission, $I_\nu$ is independent of
    $\phi$, and since there is only vacuum contributing to rays coming from 
    $\pi/2 \leq \theta \leq \pi$, 
    \begin{eqnarray}
        \mathcal{L}_\nu &=& 2\pi (4\pi r_2^2) \int_0^1 I_\nu(\mu)\, \mu\, d\mu 
        = \frac{L_{\nu,\mathrm{thin}} 4\pi r_2^2 }{2 V \alpha} \int_0^1 \{1-\exp[-\tau(\mu)]\}\mu\, d\mu ~.
    \end{eqnarray}

    We here observe that the volume-to-surface-area can be used to define a characteristic width of
    the shell, 
    \begin{eqnarray}
        \DeltaR \equiv \frac{V}{4\pi r_2^2} &=& r_2 \frac{1}{3}\left[1 - \left( \frac{r_1}{r_2} \right)^3 \right] ~, \label{eqn:dr_repr}
    \end{eqnarray}
    so the escape fraction is
    \begin{eqnarray}
        \frac{\mathcal{L}_\nu}{\mathcal{L}_\mathrm{\nu,thin}} &=& \frac{1}{2\alpha \DeltaR} 
                 \int_0^1 (1-e^{-\tau})\, \mu\, d\mu ~.
    \end{eqnarray}
    
    Here, $\mathcal{L}_\mathrm{\nu,thin}$ is the luminosity the gas would have if it were optically thin.
    In general, given the form of $\tau(\theta)$ (Equation~\ref{eqn:tau_shell}), this
    integral must be computed numerically.
    In the limit  $\tau_2\ll1$, this integral recovers 
    $\mathcal{L}_\nu = \mathcal{L}_\mathrm{\nu,thin}$
    (we note for the reader's convenience in checking this result themselves that 
    when evaluating the optically thin limit, it is helpful to define a factor $f_V = \DeltaR / r_2$
    and use $\alpha V = \tau_2 f_V$). 
    In the limit of high optical depth, the escape fraction is
    $\mathcal{L}_\nu/\mathcal{L}_\mathrm{\nu,thin} = 1/(4 \alpha \DeltaR)$.

    The shell solution has the same asymptotic behaviors as the slab approximation,
    $\mathcal{L}_\nu/\mathcal{L}_\mathrm{\nu,thin} = [1-\exp(-\tau)]/\tau$,
    if one uses $\tau = 4\alpha \DeltaR$.
    We find that the error on the escape fraction incurred by using the slab
    approximation versus numerical integration 
    depends on the thickness of the emitting region ($r_1/r_2$) and 
    $\alpha \DeltaR$ but in any case is $<10\%$. 
    The error is highest for thin shells ($r_1/r_2 \gtrsim 0.8$) and 
    near the transition between optically thick and thin regimes ($\alpha\DeltaR \sim 0.5$). 
    Therefore, we consider a slab approximation to be suitable in this work,
    and use
    \begin{equation}
        \frac{\mathcal{L}_\nu}{\mathcal{L}_\mathrm{\nu,thin}} = \frac{1-\exp(-4\alpha\DeltaR)}{4\alpha\DeltaR} \quad. 
        \label{eqn:approx_escp_noext}
    \end{equation}

    \subsection{Including Absorption by an External Medium}\label{appen:ext}

    In this scenario we have the same emitting (and self-absorbing) shell as in 
    the last case, but additionally there is absorption from an external shell that 
    extends from $r_2$ (the edge of the emission region) to $r_3$ (the edge of the CSM shell), 
    representing the as-yet-unshocked CSM. 

    The specific intensity along any path is
    \begin{eqnarray}
    I_\nu(z) = \frac{\mathcal{L}_\mathrm{\nu,thin}}{4\pi\alpha V} (1-\exp[-\tau_\mathrm{ssa}(z)]) \exp[-\tau_\mathrm{ff}(z)] \quad ,
    \end{eqnarray}
    where $z$ is the height above the equator, $\tau_\mathrm{ssa}$ is the optical depth to synchrotron self-absorption 
    (internal absorption; simply called ``$\tau$'' in the previous calculation),
    $\tau_\mathrm{ff}$ is the optical depth to free-free absorption (external absorption), 
    and all other variables are as before.
    We use $z$ rather than $\theta$ here because in terms of $z$ the integral to 
    calculate flux has the same limits with the external absorption as without.
    To maintain the definition of $\theta$ as the angle relative to the surface normal at $r_2$, 
    we define $\theta^\prime$ to be the angle relative to the surface normal at $r_3$; 
    then $z = r_2 \sin \theta = r_3 \sin \theta^\prime$, and 
    $\mu d\mu = -r_2^{-2} z dz = -r_3^{-2} z dz$. 
    For this calculation, we are evaluating the flux at $r_3$ rather than $r_2$. 
    Then
    \begin{eqnarray}
        L_\nu &=& 4\pi r_3^2 \frac{L_\mathrm{\nu,thin}}{2\alpha V}
        \int_0^{r_2} [1 - e^{-\tau_\mathrm{ssa}(z)}]e^{-\tau_\mathrm{ff}(z)} \frac{z dz}{r_3^2}\\
        \frac{L_\nu}{L_\mathrm{\nu,thin}} &=&
        4\pi \frac{1}{2\alpha V}        
        \int_0^{r_2} [1 - e^{-\tau_\mathrm{ssa}(z)}]e^{-\tau_\mathrm{ff}(z)} z dz \\
        &=& \frac{1}{2 r_2^2 \alpha \DeltaR} \int_0^{r_2} [1 - e^{-\tau_\mathrm{ssa}(z)}]e^{-\tau_\mathrm{ff}(z)} z dz \quad . \\
    \end{eqnarray}
    Defining $\zeta = z/r_2$, 
    \begin{eqnarray}
        \frac{L_\nu}{L_\mathrm{\nu,thin}} &=&
        \frac{1}{2 \alpha \DeltaR} \int_0^{1} [1 - e^{-\tau_\mathrm{ssa}(\zeta)}]e^{-\tau_\mathrm{ff}(\zeta)} \zeta d\zeta \quad . 
    \end{eqnarray}    
    This is the exact solution for the escape fraction.
    
    In this work we have approximated this result by simply accounting for external absorption with an exponential factor such that
    \begin{equation}
        \frac{\mathcal{L}_\nu}{\mathcal{L}_\mathrm{\nu,thin}} = \frac{1-\exp(-4\alpha\DeltaR)}{4\alpha\DeltaR} 
        \exp(-\alpha_\mathrm{ff} \Delta r_\mathrm{ext}) \quad. 
        \label{eqn:approx_escp_withext}
    \end{equation}
    where $\Delta r_\mathrm{ext} = r_3 - r_2$ is the radial width of the CSM and $\alpha_\mathrm{ff}$ is the 
    free-free (Bremsstrahlung) extinction coefficient, which is assumed to be constant in the
    preshock CSM. 
    
    In a case where the SSA optical depth is low, we computed the  difference
    between the result of the numerical integral and this approximation
    for various values of $\alpha_\mathrm{ff} r_3$ (optical depth) and
    $r_2/r_3$ (absorbing medium thickness). 
    We find that the error of the approximation increases as 
    this $\alpha_\mathrm{ff} r_3$ increases, 
    and that the error due to geometric effects is largest at $r_2/r_3\sim0.5$. 
    However, even for this worst case thickness, 
    the error is $\sim10\%$ at $\tau \sim 1$ 
    (approximation gives 10\% higher luminosity), 
    and 100\% at $\tau \sim 10$ (approximation gives twice the luminosity).
    The error increases by approximately a decade for each increasing decade in $\tau$, 
    but, in our view, it does not matter because the luminosity is essentially
    completely absorbed in this regime.


\bibliography{refs}
\bibliographystyle{aasjournal}



\end{document}